\documentclass[11pt,a4paper]{article}
\usepackage[utf8]{inputenc} 
\usepackage{bbold}
\usepackage{slashed}
\usepackage{a4wide}
\usepackage{amsmath}
\usepackage{cite}       
\usepackage{xcolor}
\usepackage{amssymb}
\usepackage{amsfonts}
\usepackage{booktabs}
\usepackage{makecell}
\usepackage{multirow}
\usepackage{pdflscape}
\usepackage{textcomp}
\usepackage{gensymb}
\usepackage{bigstrut}
\usepackage{array}
\usepackage{enumerate}
\usepackage{enumitem}
\usepackage[small,bf]{caption}
\usepackage{subcaption}
\usepackage{diagbox}
\usepackage{graphicx}
\usepackage{mathbbol}
\usepackage{appendix}
\usepackage{verbatim}
\usepackage[top=2.5cm,bottom=2.5cm,left=3cm,right=3cm]{geometry}
\usepackage{makecell}
\usepackage{listings}
\usepackage{mathtools}
\usepackage{dsfont}
\usepackage{braket}
\usepackage{longtable}
\usepackage{threeparttable}
\usepackage{floatpag}
\usepackage[colorlinks=true,urlcolor=myred,anchorcolor=black,citecolor=myred,filecolor=black,linkcolor=myred,menucolor=blue,linktocpage=true]{hyperref}
\usepackage[normalem]{ulem}
\usepackage{multicol}

\usepackage[capitalise]{cleveref} 
\crefname{table}{Table}{tables}

\definecolor{myred}{rgb}{0.8, 0, 0.4}
\definecolor{myblue}{cmyk}{0.8, 0.4, 0, 0.2}
\definecolor{mygreen}{rgb}{0.27, 0.64, 0.28}
\definecolor{mygray}{gray}{.95}
\definecolor{verdes}{cmyk}{0.92,0,0.59,0.4}

\DeclareMathOperator{\diag}{diag}

\DeclareMathOperator{\im}{Im}
\DeclareMathOperator{\re}{Re}

\newcommand{\id}{\mathds{1}}

\newcolumntype{x}[1]{>{\centering\arraybackslash\hspace{0pt}}p{#1}}

\makeatletter
\renewenvironment{thebibliography}[1]
{\begin{multicols}{2}[\section*{\refname}]%
		\@mkboth{\MakeUppercase\refname}{\MakeUppercase\refname}%
		\list{\@biblabel{\@arabic\c@enumiv}}%
		{\settowidth\labelwidth{\@biblabel{#1}}%
			\leftmargin\labelwidth
			\advance\leftmargin\labelsep
			\@openbib@code
			\usecounter{enumiv}%
			\let\p@enumiv\@empty
			\renewcommand\theenumiv{\@arabic\c@enumiv}}%
		\sloppy
		\clubpenalty4000
		\@clubpenalty \clubpenalty
		\widowpenalty4000%
		\sfcode`\.\@m}
	{\def\@noitemerr
		{\@latex@warning{Empty `thebibliography' environment}}%
		\endlist\end{multicols}}
\makeatother

\begin{document}

\vspace*{-15mm}
\begin{flushright}
% \revise{[@STP: update preprint number] SISSA XX/2025/FISI}
\end{flushright}
\vspace*{5mm}

\vspace{2cm}

\renewcommand*{\thefootnote}{\fnsymbol{footnote}}

\begin{center}
{\bf 
{\LARGE $S^\prime_4$ Quark Flavour Model in the Vicinity of the 
Fixed Point $\tau = i\infty$}
}
\\[8mm]

% \author{
S. T.~Petcov $^{1,2}$\footnote{Also at:
			Institute of Nuclear Research and Nuclear Energy,
			Bulgarian Academy of Sciences, 1784 Sofia, Bulgaria.}~  
		%\footnote{B's mail} 
and~ M. Tanimoto $^{3}$
		%\footnote{C's mail}
		%\footnote{D's mail}
		% \\*[20pt]
        	%\centerline
% }

\vskip 7mm

{
\begin{minipage}{\linewidth}
%\begin{center}
$^1${\it \normalsize
INFN/SISSA, Via Bonomea 265, 34136 Trieste, Italy} \\*[5pt]
$^2${\it \normalsize Kavli IPMU (WPI), UTIAS, The University of Tokyo, 
  Kashiwa, Chiba 277-8583, Japan}
\\*[5pt]
$^3${\it \normalsize
Department of Physics, Niigata University, Ikarashi 2, 
Niigata 950-2181, Japan} 
% \\*[5pt]
%\end{center}
\end{minipage}
}

\end{center}
 \vskip 7mm  
\begin{center}  
\textbf{Abstract} 
\end{center}

\small
We study in the bottom-up framework the possibility to generate the 
quark mass hierarchies without fine-tuning,
the quark mixing and CP-violation (CPV) in a flavour 
model with $S^\prime_4$ modular symmetry having minimal 
number of parameters. The model is considered in the vicinity 
of the fixed point $\tau_\text{T}= i\infty$, 
$\tau_\text{vev} \sim \tau_\text{T}$, $\tau_\text{vev}$ 
being the vacuum 
expectation value (VEV) of the modulus $\tau$, 
which allows to explain the hierarchies of the quark masses. 
The ten quark observables are described by nine real parameters. 
The CP-symmetry is broken explicitly since, as is well known, 
reproducing the observed CPV in the quark sector in the case of 
spontaneous breaking of CP-symmetry by $\tau_\text{vev}$ is highly 
problematic in the class of minimal modular quark flavour models 
(explaining the quark mass hierarchies without fine-tuning) 
of the type we consider.
We perform a statistical analysis of the model and show that 
it is phenomenologically viable and 
consistent, in particular, with the ``inclusive'' decay data 
on the  $|V_{ub}|$ and $|V_{cb}|$ elements of the CKM matrix and,
in the case of a very high scale of supersymmetry breaking, 
with the current ``average'' experimental values 
of $|V_{ub}|$ and $|V_{cb}|$.

\vspace{1em}

\renewcommand*{\thefootnote}{\arabic{footnote}}
\setcounter{footnote}{0}

%%%%%%%%%%%%%%%%%%%%%%%
%
% \clearpage
% \vfill
% \tableofcontents
% \vskip 1cm
% \hrule
% \vskip 1cm
%
%%%%%%%%%%%%%%%%%%%%%%%%%%%%%%

\newpage
%%%%%%%%%%%%%%%%%
%
\section{Introduction}
\label{Intro}
One of the problems in the otherwise very successful 
modular invariance approach to the highly challenging flavour 
problem in particle physics \cite{Feruglio:2017spp} 
(see also, e.g., \cite{deAdelhartToorop:2011re,Kobayashi:2018vbk,Penedo:2018nmg,Criado:2018thu,Kobayashi:2018scp,Chen:2019ewa,Novichkov:2018ovf,Kobayashi:2019mna,Criado:2019tzk,Ding:2019gof,Novichkov:2018nkm, Ding:2019xna,Novichkov:2018yse,Ding:2019zxk,Zhang:2019ngf,Okada:2018yrn,Kobayashi:2019rzp,deAnda:2018ecu,Okada:2019uoy,Lu:2019vgm,Chen:2021zty,Novichkov:2019sqv,Kobayashi:2019uyt,Yao:2020qyy,Ding:2021iqp,Qu:2021jdy,Liu:2021gwa,Liu:2019khw,Novichkov:2020eep,Liu:2020akv,Wang:2020lxk,Yao:2020zml,Ding:2022nzn,Ding:2022aoe,Okada:2020ukr,Okada:2020brs,Feruglio:2021dte,Wang:2021mkw,Novichkov:2021evw,Petcov:2022fjf,Abe:2023qmr,Petcov:2023vws,deMedeirosVarzielas:2023crv,King:2020qaj,Kuranaga:2021ujd,Feruglio:2023bav,Feruglio:2023mii,Ding:2024xhz,Feruglio:2023uof,Petcov:2024vph,Penedo:2024gtb,Feruglio:2024ytl,Feruglio:2025ajb,Benabou:2025viy,Ding:2024pix,Granelli:2025lds,Kobayashi:2023zzc,Ding:2024ozt,Tanimoto:2024fdl})
\footnote{A rather complete lists of articles on the 
modular-invariant 
% lepton and/or quark flavour models 
 approach to the quark and lepton flavour problems  
can be found, e.g., in \cite{deMedeirosVarzielas:2023crv,Ding:2024ozt}.
We cite here only a representative sample of the articles 
based on the bottom-up approach. 
}
is to reconcile within the bottom-up framework  
the non-fine-tuned generation of quark mass hierarchies 
with correct description of the CP-violation (CPV) in the quark sector 
%with minimal number of parameters 
when the only source of CP-symmetry breaking is the vacuum expectation 
value (VEV), $\tau_\text{vev}$, of a single modulus $\tau$ 
\cite{Petcov:2022fjf,Abe:2023qmr,Petcov:2023vws,deMedeirosVarzielas:2023crv}.
In this framework 
the quark fields transform as weighted multiplets under the action 
of the homogeneous (inhomogeneous) modular group 
\(\Gamma \equiv SL(2, \mathbb{Z})\) (\( PSL(2, \mathbb{Z})\)),
furnishing irreducible representations of the finite 
homogeneous (inhomogeneous) modular modular group of  level $N=2,3,4,5$, 
\(\Gamma^{(\prime)}_N\) - the quotient of the modular group 
and its principal congruence normal subgroup $\Gamma(N)$ 
($\overline{\Gamma}(N)$).
For a given $N = 2,3,4,5$, the finite modular group 
 \(\Gamma^{(\prime)}_N\) 
\footnote{
We recall that the groups \(\Gamma_N\) for  $N=2,3,4,5$ 
are isomorphic to the permutation groups, namely,
\(\Gamma_2 \simeq S_3\),
\(\Gamma_3 \simeq A_4\),
\(\Gamma_4 \simeq S_4\) and
\(\Gamma_5 \simeq A_5\),
while \(\Gamma^{\prime}_{2,3,4,5}\) 
are isomorphic to their respective double covers 
($S^\prime_3 \equiv S_3$).  
These non-Abelian discrete groups 
are widely used in flavour model building 
(see, e.g., \cite{Ishimori:2010au,Ishimori:2012zz,Petcov:2017ggy,Kobayashi:2022moq}).  
}
describes the flavour symmetry of the  
corresponding modular flavour model. 
The modular symmetry is broken by 
the VEV of the modulus $\tau$, which can also break 
the CP-symmetry.

 In the approach we will be interested in the present study,
the quark mass hierarchies are 
considered in the ``vicinity'' of the fixed points of the modular 
group \(\Gamma \equiv SL(2, \mathbb{Z})\) 
(\( PSL(2, \mathbb{Z})\)) \cite{Feruglio:2017spp,Novichkov:2018ovf}
\(\tau_\text{ST} = 
\omega \equiv \exp(i\,2\pi/ 3)
= -\,1/2 + i\sqrt{3}/2\) (the `left cusp') and 
 \(\tau_\text{T} = i\infty\), preserving, respectively, 
the \(\mathbb{Z}^{ST}_3\) symmetry, and, for  
the theories based on the finite modular group 
\(\Gamma^{(\prime)}_N\), $N=2,3,4,5$,   
%or \(\Gamma_N\) (see further), 
the \(\mathbb{Z}^{T}_N\) symmetry, 
$S$ and $T$ being the generators of the 
finite modular group (see, e.g., 
\cite{Feruglio:2017spp,deAdelhartToorop:2011re}
and Section 2). 
The deviation of $\tau_\text{vev}$ from the fixed  point, e.g., 
$\epsilon \sim |\tau_\text{vev} - \tau_\text{ST}|$, 
serves as a small parameter \cite{Novichkov:2021evw}. 
The elements of the quark mass matrices are expressed 
as powers of this small parameter, the exponents of which are 
uniquely determined by the transformation properties of the quark field 
components under the action of the residual symmetry group 
 \(\mathbb{Z}^{ST}_3\) or  \(\mathbb{Z}^{T}_N\), and are not 
free parameters \cite{Novichkov:2021evw}. 
The hierarchical structure of the quark mass matrices 
thus generated leads to hierarchical quark masses without the need 
to fine-tune the parameters of the corresponding models.

This approach to the quark mass hierarchies is based on the fact that 
the modular forms, which are holomorphic functions of the modulus $\tau$, 
have specific transformation properties under the action of 
the finite modular group and 
in terms of which the elements of the quark  
mass matrices are expressed in modular flavour models as a consequence 
of the modular invariance \cite{Feruglio:2017spp} 
(see also, e.g., 
\cite{Liu:2019khw,Novichkov:2020eep,Liu:2020akv,Wang:2020lxk}), 
take specific values at the fixed points 
$\tau_\text{ST}$ and $\tau_\text{T}$. Many of them have zero values 
leading to the appearance of texture zeros in the quark mass matrices.
When $\tau_\text{vev}$ deviates sufficiently slightly from the 
chosen fixed point, the texture zeros begin to be filled in
with non-zero values, whose magnitude is 
determined by the powers of the small parameter 
describing the deviation of $\tau_\text{vev}$ from the respective 
fixed point \cite{Novichkov:2021evw}.
Thus, hierarchical quark mass matrices are generated leading to a 
non-fine-tuned  generation of the quark mass hierarchies. 

Using this approach a phenomenologically viable non-fine-tuned 
model of charged lepton mass hierarchies and neutrino mixing 
based on the $S_4$ modular symmetry considered in the vicinity of the 
fixed point $\tau_\text{ST}$ was first constructed in \cite{Novichkov:2021evw}.
When implementing in the quark sector,    
the approach described above faces, in the case of minimal models 
\footnote{By ``minimal'' we mean, in particular, models 
which do not include additional flavon fields.
}
with  CP-conserving constant parameters and only  
one modulus $\tau$ whose VEV is the only source of CP-symmetry breaking,  
the problem of reconciling the relatively small values of the 
non-diagonal elements of the quark mass matrices, needed to generate 
the quark mass hierarchies, with reproducing the relatively large 
value of the CPV phase in the CKM mixing matrix 
(see, e.g., 
\cite{Petcov:2022fjf,Abe:2023qmr,Petcov:2023vws,deMedeirosVarzielas:2023crv}). 
To avoid this problem modular quark flavour models 
with explicit breaking of the CP-symmetry have been proposed.
Typically in these models the number of parameters 
is larger than the number of the quark observables. 
In \cite{Abe:2023qmr}, for example, such a model, 
 based on  
$S^\prime_4$  modular symmetry considered 
in the vicinity of the fixed  point $\tau_\text{T}= i\infty$ 
having 14  real parameters and 2 phases besides 
$\tau_\text{vev}$,
has been constructed in the quark sector, 
and it was shown that it can successfully 
reproduce the observed quark masses and CKM mixing matrix.
%
% Model with 12 parameters:
%\cite{Abe:2023ilq}
% \bibitem{Abe:2023ilq}
% Y.~Abe, T.~Higaki, J.~Kawamura and T.~Kobayashi,
% %``Quark masses and CKM hierarchies from $S_4'$ modular flavor symmetry,''
% Eur. Phys. J. C \textbf{83} (2023) no.12, 1140
% doi:10.1140/epjc/s10052-023-12303-2
% [arXiv:2301.07439 [hep-ph]].

In the present article we study the quark mass hierarchies, 
the CKM mixing angles and CPV phase close to the 
fixed point \(\tau_\text{T} = i\infty\) in a ``minimal'' 
quark flavour model based on the $S^\prime_4$ symmetry.
The residual symmetry at  \(\tau_\text{T} = i\infty\) 
in this case is \(\mathbb{Z}_4\). 
The small parameter describing the deviation of 
$\tau_\text{vev}$ from the fixed point is \cite{Novichkov:2021evw}
$|q_4| = \exp(-\,2\pi\,\im [\tau_\text{vev}]/4) \ll 1$. 
The ten quark observables are described in the model by nine real parameters,
 namely,
six real constants, one complex constant and $\im [\tau_\text{vev}]$.
The CP-symmetry is broken explicitly since, as we have already discussed, 
reproducing the observed CPV in the quark sector in the case of 
spontaneous breaking of CP-symmetry by $\tau_\text{vev}$ is highly 
problematic in the class of minimal modular quark flavour models 
(explaining the quark mass hierarchies without fine-tuning) 
of the type we consider. 
Thus, we do not impose the gCP-symmetry  \cite{Novichkov:2019sqv}.
The CP-symmetry breaking is 
provided by the only complex constant that can be present 
naturally, i.e., without additional assumptions, in the model 
when the gCP-symmetry does not hold. We solve the model completely 
analytically, i.e., we derive 
sufficiently precise analytic expressions for all ten observables  
in terms of the constant parameters of the model and $|q_4|$. 
These expressions turned out to be surprisingly simple. 
We perform statistical analyses of the model 
assuming relatively low SUSY breaking scale $M_{\rm SUSY} = 10$ TeV and
using, in particular, the ``inclusive'' and ``exclusive'' decay data on,  
and the ``average'' experimental values of,  
the $|V_{ub}|$ and $|V_{cb}|$ elements of the CKM matrix,
as well as in the case of very high $M_{\rm SUSY}$   
and the ``average'' experimental values of 
the $|V_{ub}|$ and $|V_{cb}|$.

 After introducing the modular symmetry briefly in 
Section~\ref{sec:Modular}, we present the general framework of 
generating the mass hierarchy close to $\tau=i\infty$ without fine-tuning 
in Section~\ref{sec:theory}.
In Sections \ref{sec:model}
we construct a ``minimal''  $S^\prime_4$ modular invariant
 quark flavour model, 
describing  how one can naturally generate 
hierarchical quark mass patterns 
in the vicinity of the fixed point \(\tau_\text{T} = i\infty\)
and obtain analytic expressions for the 
quark mass matrices and the quark mass ratios.
This allows us to derive analytic expressions for 
the mixing matrices originating from the 
down-type quark and up-type quark sectors in 
Sections \ref{sec:dmixing} and
\ref{sec:umixing}, and for  
the CKM matrix in Section \ref{sec:CKM}.
After summarizing the input quark data, 
we perform numerical analysis of the 
description of the quark mass hierarchies
and CKM parameters by the constructed ``minimal'' model 
in Section \ref{sec:numerics}. 
We summarize our results  in Section~\ref{sec:summary}.

%%%%%%%%%%%%%%%%%%%%%%%%%%%
%
\section{Modular symmetry and its Residual Symmetries}
\label{sec:Modular}
%
%%%%%%%%%%%%%%%%%%%%%%%%%%%%%%%%%%%%%%%%%%
%
The modular invariance approach to flavour is usually considered 
within the SUSY framework. This severely constraints the form of the 
fermion Yukawa couplings and mass matrices, limiting the number of 
possible terms, and ensures that the results are stable against possible 
corrections. 

The elements of the (homogeneous) modular group 
\(\Gamma \equiv SL(2, \mathbb{Z})\)
describing the modular symmetry are  
$2\times 2$ matrices $\gamma$ with integer elements and 
determinant euqal to 1:
%%%%%%%%%%%%%%%%%%%%%%%%%%
\begin{equation}
\label{eq:gamma}
\gamma =
\begin{pmatrix}
a & b \\ c & d
\end{pmatrix}
\in \Gamma\, \quad ad - bc = 1\,. 
% \tau \to \gamma \tau = \frac{a\tau + b}{c\tau + d} \,,
\end{equation}
%%%%%%%%%%%%%%%%%%%%%%%%%%%
%
The group generators are:
%%%%%%%%%%%%%%%%%%%%%%%%%%%%%
\begin{equation}
\label{eq:STR_def}
S =
\begin{pmatrix}
0 & 1 \\ -1 & 0
\end{pmatrix}
\,, \quad
T =
\begin{pmatrix}
1 & 1 \\ 0 & 1
\end{pmatrix}
\,, \quad
R =
\begin{pmatrix}
-1 & 0 \\ 0 & -1
\end{pmatrix}\,,
\end{equation}
%
%%%%%%%%%%%%%%%%%%%%%%%%%%%%%
%
satisfying \(S^2 = R\), \((ST)^3 = R^2 = \id\), and \(RT = TR\).

The matter superfields transform under the action of the 
(homogeneous) modular group 
\(\Gamma \equiv SL(2, \mathbb{Z})\) as weighted multiplets 
~\cite{Feruglio:2017spp,Ferrara:1989bc,Ferrara:1989qb}:
%%%%%%%%%%%%%%%%%%%%%%%%%%%%%%%%
\begin{equation}
\label{eq:psi_mod_trans0}
\psi_i \to (c\tau + d)^{-k} \, \rho_{ij}(\gamma) \, \psi_j \,,
\end{equation}
%%%%%%%%%%%%%%%%%%%%%%%%%%%%%%%%%%%%%%%%%%%
%
where \(k \in \mathbb{Z}\) is the so-called modular weight%
\footnote{While we restrict ourselves to integer \(k\), it is also possible 
for weights to be fractional 
\cite{Dixon:1989fj,Ibanez:1992hc,Olguin-Trejo:2017zav,Nilles:2020nnc}.
}
and \(\rho(\gamma)\) is a unitary representation of~\(\Gamma\).
In using modular symmetry as a flavour symmetry 
one assumes that \(\rho(\gamma) = \id\) 
for elements \(\gamma\) of the principal congruence normal 
subgroup $\Gamma(N)$ of \(\Gamma\),
% $SL(2, \mathbb{Z})$,  
\(N \geq 2\) being a natural number called ``level'':
%%%%%%%%%%%%%%%%%%%%%%%%%%%%%%%%%%%
\begin{equation}
\label{eq:congr_subgr}
\Gamma(N) \equiv
\left\{
\begin{pmatrix}
a & b \\ c & d
\end{pmatrix}
\in SL(2, \mathbb{Z}), \,
\begin{pmatrix}
a & b \\ c & d
\end{pmatrix}
\equiv
\begin{pmatrix}
1 & 0 \\ 0 & 1
\end{pmatrix}
(\text{mod } N)
\right\}\,.
\end{equation}
%%%%%%%%%%%%%%%%%%%%%%%%%%%%%%%%%%
%
Thus, \(\rho\) is effectively a representation of the (homogeneous) 
finite modular group of level $N$,   
\(\Gamma^\prime_N \equiv \Gamma\,\big/\, \Gamma(N) \simeq SL(2, \mathbb{Z}_N)\). 
For level \(N\leq 5\), this group admits the presentation:
%%%%%%%%%%%%%%%%%%%%%%%%
\begin{equation}
\label{eq:hom_fin_mod_group_pres}
\Gamma'_N = \left\langle S, \, T, \, R \mid S^2 = R, \, (ST)^3 = \id, \, R^2 = \id, \, RT = TR, \, T^N = \id \right\rangle\,.
\end{equation}
%%%%%%%%%%%%%%%%%%%%%%%%%%%%%%%
%

In the considered bottom-up framework  one introduces a chiral superfield, 
the modulus~\(\tau\), transforming non-trivially 
under the action of the  modular group,  
% \(\Gamma \equiv SL(2, \mathbb{Z})\). 
% \(\Gamma\), 
the elements \(\gamma\) of 
% the modular group 
\(\Gamma\) acting on \(\tau\) via the fractional linear transformation:
%%%%%%%%%%%%%%%%%%%%%%%%%%
\begin{equation}
\label{eq:tau_mod_trans}
\gamma =
\begin{pmatrix}
a & b \\ c & d
\end{pmatrix}
\in \Gamma : \quad
\tau \to \gamma \tau = \frac{a\tau + b}{c\tau + d}\,.
\end{equation}
%%%%%%%%%%%%%%%%%%%%%%%%%%%
%
The  modulus~\(\tau\) acquires a VEV, which is restricted to the upper 
half of the complex plane.
It plays the role of a spurion and
parametrises the breaking of the modular (flavour) symmetry. 
Additional flavon fields are not needed to break the symmetry, 
and we do not introduce them in our study.
The VEV of \(\tau\) can also be the only 
source of CP-symmetry breaking in a modular flavour model.
 
Since~\(\tau\) does not transform under the \(R\) 
generator of \(\Gamma\), a \(\mathbb{Z}_2^R\) symmetry is preserved 
in such scenarios \cite{Novichkov:2020eep}.
If also matter fields transform trivially under \(R\), one may identify 
the matrices \(\gamma\) and \(-\gamma\), thereby restricting oneself to the 
inhomogeneous modular group~
\(\overline{\Gamma} \equiv PSL(2, \mathbb{Z}) 
\equiv SL(2, \mathbb{Z}) \, / \, \mathbb{Z}_2^R\). 
In such a case, \(\rho\) is effectively a representation of a smaller 
(inhomogeneous) finite modular group of level $N$: 
\(\Gamma_N \equiv \Gamma \, \big/ \left\langle \, \Gamma(N) \cup \mathbb{Z}_2^R \, \right\rangle\). 
For \(N\leq 5\), this group admits the presentation
%%%%%%%%%%%%%%%%%%%%%%%%%
\begin{equation}
\label{eq:inhom_fin_mod_group_pres}
\Gamma_N = \left\langle S, \, T \mid S^2 = \id, \, (ST)^3 = \id, \, T^N = \id \right\rangle \,.
\end{equation}
%%%%%%%%%%%%%%%%%%%%%%%%%%%%%%%%%5
%
In general, however, \(R\)-odd fields may be present in the theory and 
\(\Gamma\) and \(\Gamma_N'\) are then the relevant symmetry groups.
For $N=2,3,4,5$,  $\Gamma_N$, as is well known, 
is isomorphic to the non-Abelian discrete symmetry groups
$S_3$, $A_4$, $S_4$, $A_5$ and  $\Gamma_N'$ is isomorphic to their 
respective double covers.

Finally, to understand how modular symmetry may constrain the Yukawa couplings 
and mass structures of a model in a predictive way, we turn to the 
Lagrangian, which for an \(\mathcal{N} = 1\) global supersymmetric theory 
is given by:
%%%%%%%%%%%%%%%%%%%%%%%%%%%%%
\begin{equation}
\mathcal{L} = \int \text{d}^2 \theta \, \text{d}^2 \bar{\theta} \, K(\tau,\bar{\tau}, \psi_I, \bar{\psi}_I)
+ \left[ \, \int \text{d}^2 \theta \, W(\tau,\psi_I) + \text{h.c.} \right] \,.
\end{equation}
%%%%%%%%%%%%%%%%%%%%%%%%%%%%%%
%
Here \(K\) and \(W\) 
are the K\"ahler potential and the superpotential, respectively. 
The superpotential \(W\) 
can be expanded in powers of matter superfields \(\psi_I\),
%%%%%%%%%%%%%%%%%%%%%%%%
\begin{equation}
\label{eq:W_series}
W(\tau, \psi_I) = \sum \left( \vphantom{\sum} Y_{I_1 \ldots I_n}(\tau) \, \psi_{I_1} \ldots \psi_{I_n} \right)_{\mathbf{1}} \,,
\end{equation}
%%%%%%%%%%%%%%%%%%%%%%%%
% 
where one has summed over all possible field combinations and 
independent singlets of the finite modular group.
 By assuming that the fields $\psi_{I_i}$ furnish unitary irreducible 
representation $\rho_{I_i}$ of a finite modular group of level $N$,
$\Gamma_N^{(\prime)}$, and carry weights $k_{I_i}$, and requiring  
that the superpotential is invariant
under the modular transformations, one finds that the field couplings 
\(Y_{I_1 \ldots I_n}(\tau)\) have to be modular forms of level \(N\).
These are severely constrained holomorphic 
functions of~\(\tau\), which under modular transformations obey
%%%%%%%%%%%%%%%%%%%%%%%%%%%%%%%%
\begin{equation}
\label{eq:Y_mod_trans}
Y_{I_1 \ldots I_n}(\tau) \,\xrightarrow{\gamma}\, Y_{I_1 \ldots I_n}(\gamma \tau) = (c\tau + d)^{k} \rho_Y(\gamma) \,Y_{I_1 \ldots I_n}(\tau) \,,
\end{equation}
%%%%%%%%%%%%%%%%%%%%%%%%%%%%%%%
%
 where $k$ is the modular form weight.
The modular invariance implies that
the modular form \(Y_{I_1 \ldots I_n}(\tau)\) should 
carry weight \(k = k_{I_1} + \ldots + k_{I_n}\) and should
furnish unitary irreducible representations \(\rho_{Y}\) of the finite 
modular group $\Gamma_N^{(\prime)}$ such that 
\(\rho_Y \,\otimes\, \rho_{I_1} \,\otimes \ldots \otimes\, \rho_{I_n} \supset \mathbf{1}\).
Non-trivial modular forms of a given level exist only for  
% integer positive \(k)\
\(k \in \mathbb{N}\), span finite-dimensional linear spaces
\(\mathcal{M}_{k}(\Gamma(N))\), and can be arranged into multiplets 
of \(\Gamma^{(\prime)}_N\) 
\cite{Feruglio:2017spp,Kobayashi:2018scp,Penedo:2018nmg,Novichkov:2018nkm,Liu:2019khw,Novichkov:2020eep,Wang:2020lxk}.

The breakdown of modular symmetry is parametrised by the VEV of the modulus 
and there is no value of \(\tau\) which preserves the full symmetry. 
Nevertheless, at certain so-called symmetric points 
\(\tau = \tau_\text{sym}\) (called  also ``fixed points''),
the modular group is only partially broken, with the unbroken generators 
giving rise to residual symmetries.
As we have indicated, the \(R\) generator is unbroken for any 
value of \(\tau\), so that a \(\mathbb{Z}_2^R\) symmetry is always preserved.
There are only three inequivalent symmetric points (in the fundamental domain
of the modular group), namely~
\cite{Novichkov:2018ovf}:
%%%%%%%%%%%%%%%%%%%%%%%%%%%%%%
{\bf 
\begin{itemize}
	\item \(\tau_\text{sym} \equiv \tau_\text{T} =
i \infty\), invariant under \(T\), 
preserving \(\mathbb{Z}_N^T \times \mathbb{Z}_2^R\)\,,
	\item \(\tau_\text{sym} \equiv \tau_\text{S} = i\), 
invariant under \(S\), preserving \(\mathbb{Z}_4^S \) 
(recall that \(S^2 = R\))\,,
	\item \(\tau_\text{sym} \equiv \tau_\text{ST} = \omega 
\equiv \exp (2\pi i / 3)\), invariant under 
\(ST\), preserving \(\mathbb{Z}_3^{ST} \times \mathbb{Z}_2^R\)\,.
\end{itemize}
}
%%%%%%%%%%%%%%%%%%%%%%%%%%%
%

%%%%%%%%%%%%%%%%%%%%%%%
%
\section{Mass hierarchy without fine-tuning close to $\tau_T=i\infty$}
\label{sec:theory}
%
%%%%%%%%%%%%%%%%%%%%%%%

In theories where modular invariance 
is broken only by the VEV of modulus $\tau$, 
the fermion mass matrices (in the limit of unbroken SUSY)
are expressed in terms of modular forms of a given level $N$
and a limited number of coupling constants in the superpotential 
(see, e.g.,  
\cite{Feruglio:2017spp,Kobayashi:2018scp,Penedo:2018nmg,Criado:2018thu,Novichkov:2018ovf,Kobayashi:2019mna,Criado:2019tzk,Ding:2019gof,Novichkov:2018nkm}).
The flavour structure of the mass matrices 
is determined by the properties of the respective modular forms 
at the value of the VEV of \(\tau\). It  can be severely constrained 
at the points of residual symmetries \(\tau = \tau_\text{sym}\) 
which typically enforce the presence of multiple zero entries 
in the mass matrices due to zero values of the corresponding modular 
form components. As \(\tau\)  moves away from its symmetric value, 
these entries will generically become non-zero 
but are suppressed and thus a flavour structure arises. 

This approach to fermion (charged lepton and quark) mass hierarchies was 
explored in, e.g., \cite{Feruglio:2021dte,Novichkov:2021evw,Petcov:2022fjf}.
We present below a more detailed discussion of the approach   
following \cite{Novichkov:2021evw}.
If $\epsilon$ parameterises  the deviation of  \(\tau\)  from a
given symmetric point \(\tau_\text{sym}\), 
$|\epsilon| \ll 1$, the degree of suppression 
of the (residual-)symmetry-breaking entries 
is determined by $|\epsilon|^l$, 
where $l > 0$ is an integer. 
% \cite{Novichkov:2021evw}. 
The integer constant $l$ is not another 
free parameter of the theory. The values $l$ can take 
depend on the symmetric point, i.e., on the 
residual symmetry. As was shown in \cite{Novichkov:2021evw}, 
i) for $\tau_\text{S} = i$, $l=0,1$,
ii) for $\tau_\text{ST} = \omega$, $l=0,1,2$, 
and iii) for $\tau_{T} = i\infty$ 
of interest and  $\Gamma_N$ ($\Gamma^\prime_N$) finite modular group, 
$N\leq 5$, $l$ can take values $l = 0,1,...,N-1$. 
Thus, for  $\Gamma^\prime_4 \simeq S^\prime_4$ and  $\tau_{T} = i\infty$, 
$l = 0,1,2,3$. For a specific would-be-zero element of the 
fermion mass matrix the value of $l$ is uniquely 
determined by the residual symmetry  
and by the transformation properties under the 
residual symmetry group of the fermion fields, 
associated with that entry \cite{Novichkov:2021evw}.

Indeed, consider the  modular-invariant bilinear
%%%%%%%%%%%%%%%%%%%%%%%%%%
\begin{equation}
\label{eq:bilinear}
\psi^c_i \, M(\tau)_{ij}\, \psi_j \,,
\end{equation}
%%%%%%%%%%%%%%%%%%%%%%%%%%%%%%
%
where the superfields \(\psi\) and \(\psi^c\) transform under the 
modular group as%
\footnote{Note that in the case of a Dirac bilinear, \(\psi\) and \(\psi^c\) 
are independent fields, so in general \(k^c \neq k\) and 
\(\rho^c \neq \rho, \rho^{*}\).}
%%%%%%%%%%%%%%%%%%%%%%%%%
\begin{equation}
\label{eq:psi_mod_trans}
\begin{split}
\psi \,&\xrightarrow{\gamma}\, (c \tau + d)^{-k} \rho(\gamma) \,\psi \,, \\
\psi^c \,&\xrightarrow{\gamma}\, (c \tau + d)^{-k^c} \rho^c(\gamma)\, \psi^c \,,
\end{split}
\end{equation}
%%%%%%%%%%%%%%%%%%%%%%%%%%%%%%%%
%
so that each \(M(\tau)_{ij}\) is a modular form of level \(N\) and 
weight \(K \equiv k+k^c\).
Modular invariance requires \(M(\tau)\) to transform as 
%%%%%%%%%%%%%%%%%%%%%%%%%%%%%%
\begin{equation}
\label{eq:mass_matrix_mod_trans}
M(\tau)\, \xrightarrow{\gamma}\, M(\gamma \tau) 
= (c \tau + d)^K \rho^c(\gamma)^{*} M(\tau) \rho(\gamma)^{\dagger} \,.
\end{equation}
%%%%%%%%%%%%%%%%%%%%%%%%%%%%%%
%
Taking \(\tau\) to be close to a symmetric point
\footnote{
 Not to burden the notations we use $\tau$ instead of 
$\tau_\text{vev}$ in what follows.
}, and setting \(\gamma\) 
to the residual symmetry generator, one can use this transformation 
rule to constrain the form of the mass matrix \(M(\tau)\) 
\cite{Novichkov:2021evw}. 

At \(\tau_\text{T} = i\infty\) of interest we have  
$\mathbb{Z}^T_N$ residual symmetry group generated by the $T$ generator of 
$\Gamma_{\rm N}$ ($\Gamma^\prime_{\rm N}$).
Consider the $T$-diagonal representation basis for 
group generators in which \(\rho^{(c)}(T) =\diag (\rho^{(c)}_i)\), 
and assume that \(\tau\) is ``close'' to
\(\tau_\text{T} = i\infty\), i.e., that $\im[\tau]$ 
is sufficiently large. 
By setting \(\gamma = T\) in Eq.\,\eqref{eq:mass_matrix_mod_trans}, 
one finds
%%%%%%%%%%%%%%%%%%%%%%%%%%%%%%%%%%
\begin{equation}
\label{eq:mass_matrix_T_trans}
M_{ij}(T \tau) = \left( \rho^c_i\, \rho_j \right)^{*} M_{ij}(\tau) \,.
\end{equation}
%%%%%%%%%%%%%%%%%%%%%%%%%%%%%%
%
It is now convenient to treat the \(M_{ij}\) as functions of 
$\hat q \equiv \exp(i2\pi \tau/N)$, so that 
%%%%%%%%%%%%%%%%%%%%%%%%%%%%%%%
\begin{equation}
\epsilon  \equiv |\hat q| = e^{-\,2\pi \im[\tau]/N} \, 
\label{eq:q}
\end{equation}
%%%%%%%%%%%%%%%%%%%%%%%%%%%%%%
%
parametrises the deviation of \(\tau\) 
from the symmetric point \cite{Novichkov:2021evw}. 
Note that the entries \(M_{ij}(\hat q)\) depend analytically 
on \(\hat q\) and that \(\hat q \xrightarrow{T} \xi \hat q\), 
with $\xi = \exp(i\,2\pi/N)$.
Thus, in terms of \(\hat q\), 
Eq.\,\eqref{eq:mass_matrix_T_trans} reads:
%%%%%%%%%%%%%%%%%%%%%%%%%%%%%%%%%
\begin{equation}
M_{ij}(\xi \hat q) =  (\rho^c_i\, \rho_j)^{*} M_{ij}(\hat q)\,.
\end{equation}
%%%%%%%%%%%%%%%%%%%%%%%%%%%%
%
Expanding both sides in powers of \(\hat q\), one finds:
%%%%%%%%%%%%%%%%%%%%%%%%%%%%%
\begin{equation}
\label{eq:expansion_ST}
\xi^{n} M_{ij}^{(n)}(0) = 
(\rho^c_i \, \rho_j)^{*} M_{ij}^{(n)}(0)\,,
\end{equation}
%%%%%%%%%%%%%%%%%%%%%%%%%%%%%%%%%%%%%
%
where \(M_{ij}^{(n)}\) denotes the \(n\)-th derivative of 
\(M_{ij}\) with respect to \(\hat q\).
It follows that \(M^{(n)}_{ij}(0)\) can only be non-zero for values 
of $n$ such that $(\rho^c_i\, \rho_j)^{*} = \xi^n$. In particular,
the entry \(M_{ij} = M^{(0)}_{ij}(0)\)
is only allowed to be non-zero and be \(\mathcal{O}(1)\) 
if {\(\rho^c_i \rho_j = 1\)}. More generally, 
if $(\rho^c_i\, \rho_j)^{*} = \xi^\ell$ with  \(\ell=0,1,2,...,N-1\),
%%%%%%%%%%%%%%%%%%%%%%%%%%%%%%%%%%%%%%%
\begin{equation}
M_{ij}(q) = a_0\,\hat q^{\ell} + a_1\,\hat q^{\ell + N} 
+  a_2\,\hat q^{\ell + 2N} + ...\,,  
\label{eq:Mq}
\end{equation}
%%%%%%%%%%%%%%%%%%%%%%%%%
%
in the vicinity of the symmetric point. 
It crucially follows that the entry $M_{ij}$ is expected to be 
${\cal O}(\epsilon^{\ell})$ whenever $\im[\tau]$ is 
sufficiently large 
\footnote{In practice, values of $\im [\tau] \cong (2.5 - 3.0)$ 
prove to be sufficiently large (see \cite{Novichkov:2021evw} and further).}.
The power $\ell$ is uniquely determined by how the 
$\Gamma_{\rm N}$ ($\Gamma^\prime_{\rm N}$) representations 
of $\psi^c_i$ and $\psi_j$ decompose under the residual symmetry 
group $\mathbb{Z}^T_N$ \cite{Novichkov:2021evw}.
In the case of $S^\prime_4$, as we have already indicated, 
$\ell$ can take values $\ell = 0,1,2,3$.

The discussed result allows, in principle, to obtain fermion mass hierarchies 
without fine-tuning. 
 
%%%%%%%%%%%%%%%%%%%%%%%%
%
\section{The Quark Flavour Model} 
\label{sec:model}
%
%%%%%%%%%%%%%%%%%%%%%%%%%%%

We assume that the quark doublet superfields $Q$ furnish 
the triplet representation $\mathbf{3}$
of the finite modular group $S^\prime_4$ and have a weight 
$k_Q$, while the RH up-type quark  superfields $u^c$, $c^c$ and $t^c$ 
and  RH down-type quark  
superfields $d^c$, $s^c$ and $b^c$ 
furnish the singlet representations 
$\mathbf{1}$, $\mathbf{\hat{1}}$, $\mathbf{\hat{1}}$
and $\mathbf{1}$, $\mathbf{1}$, $\mathbf{\hat{1}}$, respectively, 
and have weights $k_{u^c}$, $k_{c^c}$,$k_{t^c}$ and 
$k_{d^c}$, $k_{s^c}$,$k_{b^c}$.
We assume further that $k_Q + k_{u^c} = 4$, 
$k_Q + k_{c^c} = 3$, $k_Q + k_{t^c} = 5$,
$k_Q + k_{d^c} = 6$, 
$k_Q + k_{s^c} = 4$ and $k_Q + k_{b^c} = 7$.

%%%%%%%%%%%%%%%%%%%%%%%%%%%%%%%
%
\subsection{Up-type Quarks}
%
%%%%%%%%%%%%%%%%%%%%%%%%%%%%%%%

Under the made assumptions, the 
up-type quark mass matrix reads in the 
 right-left (R-L) convention:
%%%%%%%%%%%%%%%%%%%%%%%%%%%%%
\begin{equation}
    M_u =  \dfrac{v_u}{\sqrt{3}}
\begin{pmatrix}
\hat\alpha_u & 0  & 0 \\[2mm]
0 & \hat\beta_u & 0  \\[2mm]
0 & 0&  \hat\gamma_u  
\end{pmatrix}\,
\begin{pmatrix}
\left(Y^{(4)}_{\mathbf{3}}\right)_1 & \left(Y^{(4)}_{\mathbf{3}}\right)_3 & 
\left(Y^{(4)}_{\mathbf{3}}\right)_2
\\[2mm]
\left(Y^{(3)}_{\mathbf{\hat{3}'}}\right)_1 & \left(Y^{(3)}_{\mathbf{\hat{3}'}}\right)_3 &
\left(Y^{(3)}_{\mathbf{\hat{3}'}}\right)_2
\\[2mm]
\left(Y^{(5)}_{\mathbf{\hat{3}'}}\right)_1 & \left(Y^{(5)}_{\mathbf{\hat{3}'}}\right)_3 
& \left(Y^{(5)}_{\mathbf{\hat{3}'}}\right)_2
\end{pmatrix}\,.
\label{eq:MuY}
\end{equation}
%%%%%%%%%%%%%%%%%%%%%%%%%%%%%%%
%
Here $v_u = v\sin\beta$ is the VEV of the SUSY Higgs field $H_u$
having weak hypercharge (+ 1), 
$v=174$ GeV, $\beta$ is an angle parameter playing important role 
in SUSY theories, and 
 $Y^{(4)}_{\mathbf{3}}$, $Y^{(3)}_{\mathbf{\hat{3}'}}$
and $Y^{(5)}_{\mathbf{\hat{3}'}}$ are weight 4, 3 and 5 
level 4 modular forms furnishing 
the triplet representations 
$\mathbf{3}$, $\mathbf{\hat{3}'}$ and $\mathbf{\hat{3}'}$
of $S^\prime_4$.
The constants $\hat\alpha_u$, $\hat\beta_u$ and $\hat\gamma_u$
include the rescaling factors of the matter fields 
needed to cast their kinetic terms in canonical form: 
%%%%%%%%%%%%%%%%%%%%%
\begin{eqnarray}
\nonumber
 & \hat\alpha_u = 
\alpha_u\, (2 \im[\tau])^{\frac{4}{2}}\,,\\[0.25cm]
\nonumber
& \hat\beta_u = \beta_u\,(2 \im [\tau])^{\frac{3}{2}}\,,\\[0.25cm]
&\hat\gamma_u = \gamma_u\,(2\im[\tau])^{\frac{5}{2}}\,.
%
% \hat\alpha_{3,i} \rightarrow \alpha_{3,i}=
% \hat\alpha_{3,i}\, (2Im\tau))^{\frac{5}{2}}\,,~i=1,2\,.
\label{eq:hatuconst}
\end{eqnarray}
%%%%%%%%%%%%%%%%%%%%%%%%%%%%%%
%
The constant 
parameters of the model $\alpha_u$, $\beta_u$ and $\gamma_u$ 
can be taken to be real without loss of generality 
(even without imposing the gCP-symmetry on the 
superpotential/Lagrangian) of the theory.

Using the methods developed in \cite{Novichkov:2021evw}, 
one finds the following limiting structure of $M_u$ 
''close'' to the fixed point $\tau_\text{T} = i\infty$, i.e., for 
relatively large values of $\im[\tau]$:
%%%%%%%%%%%%%%%%%%%%%%%%%%%
\begin{equation}
M_{u} \sim \begin{pmatrix}
 \epsilon^3 & \epsilon^2 & 1 \\
 \epsilon^2& \epsilon & \epsilon^3 \\
 \epsilon^3 & \epsilon^2 & 1
\end{pmatrix}\,,
\end{equation}
%%%%%%%%%%%%%%%%%%%%%%
%
where $\epsilon = |q_4| =  e^{-\,\frac{\pi}{2} \im[\tau]} \ll 1$.
% $\exp(-\,\pi Im\tau/2)$.
It leads to a hierarchical 
up-type quark mass spectrum of the type
{\bf $(m_t:m_c:m_u)\sim m_t(1:\epsilon:\epsilon^3)$.}

Using the $q-$expansions of the 
modular forms present in Eq. (\ref{eq:MuY}) 
in the $T$-diagonal basis, 
it is possible to obtain to leading order 
in $|q_4|$ analytic expressions for the up-quark
mass matrix. 
Working with the normalisation of the relevant modular
forms used in\cite{Novichkov:2020eep} 
and keeping terms not smaller than $|q_4|^4$
we find:
%%%%%%%%%%%%%%%%%%%%%%%%%%%
\begin{equation}
M_{u} \cong \dfrac{v_u}{\sqrt{3}}\,
\begin{pmatrix}
\hat\alpha_u & 0  & 0 \\[2mm]
0 & \hat\beta_u & 0  \\[2mm]
0 & 0&  \hat\gamma_u  
\end{pmatrix}\,
\begin{pmatrix}
 6\,q^2_4 & -\,\dfrac{3}{\sqrt{2}}\,q_4 & \dfrac{12}{\sqrt{2}}\,q^3_4 \\
 -\,16\sqrt{2}\,q^3_4 & -\,6\,q^2_4 & \dfrac{1}{2}(1 + 60\,q^4_4) \\
 16\,q^3_4 & \dfrac{3}{\sqrt{2}}\,q^2_4 & \dfrac{1}{4\sqrt{2}}(1-204\,q^4_4)
\end{pmatrix} + O(10^{-5})\,,
\label{eq:Muq4}
\end{equation}
%%%%%%%%%%%%%%%%%%%%%%
%
where $q_4 = \exp({i\pi \tau/2})$.
From the derived analytic expression for $M_{u}$ 
it is not difficult to derive also analytic expressions for the 
up-quark masses. We get: 
%%%%%%%%%%%%%%%%%%%%%%%%%%%
\begin{eqnarray}
\nonumber
 m_t&=& \dfrac{v_d}{\sqrt{3}}\,\dfrac{1}{2}\,
\sqrt{ |\hat\beta_u|^2 + \frac{1}{8} |\hat\gamma_u|^2}\,,\\[0.25cm]
\nonumber
 m_c &=& \dfrac{v_d}{\sqrt{3}}\,\dfrac{3}{\sqrt{2}}\,
|q_4|\,|\hat\alpha_u|\,,\\[0.25cm]
 m_u &=&  \dfrac{v_d}{\sqrt{3}}\, |q_4|^3\, 
\dfrac{36 |\hat\beta_u \hat\gamma_u|}
{\sqrt{|\hat\beta_u|^2 + \frac{1}{8} |\hat\gamma_u|^2}}\,.
\label{eq:mumcmt}
\end{eqnarray}
%%%%%%%%%%%%%%%%%%%%%%%
%
For the up-quark mass ratios we will use the values 
at the GUT scale $2\times 10^{16}$ GeV 
obtained taking into account the RG running effects 
for $\tan\beta = 10$ and $M_{\rm SUSY} = 10$ TeV \cite{Antusch:2025fpm}: 
%%%%%%%%%%%%%%%%%%%%%%%%%5
\begin{eqnarray}
\label{eq:mumt}
& \dfrac{m_u}{m_t} = (5.67 \pm 0.229) \times 10^{-6}\,,\\[0.25cm]
\label{eq:mcmt}
& \dfrac{m_c}{m_t} = (2.87 \pm 0.105) \times 10^{-3}\,.
\end{eqnarray}
%%%%%%%%%%%%%%%%%%%%%
%

The value of the ratio of the constants 
$|\hat\gamma_u|$ and $|\hat\beta_u|$
can be determined using the value of
the quark mass ratio $m_u/m_t$. 
If we denote by $X =|\hat\gamma_u|/|\hat\beta_u|$ and 
set $|q_4| = A\times 10^{-2}$ and $m_u/m_t = 5.400\,F_{ut}\times 10^{-6}$,
the ratio $X$ satisfies the following quadratic equation:
$72\,A^3\,X = 5.400\,F_{ut} (1 + X^2/8)$.
The two solutions are given approximately by:
%%%%%%%%%%%%%%%%%%%%%%%%%5
\begin{eqnarray}
\nonumber
& X_1 \equiv \left (\dfrac{\hat\gamma_u}{\hat\beta_u}\right)_1 
\cong 0.075 \dfrac{F_{ut}}{A^3}\,,\\[0.25cm]
& X_2 \equiv \left (\dfrac{\hat\gamma_u}{\hat\beta_u}\right)_2
\cong  \dfrac{32}{0.3} \dfrac{A^3}{F_{ut}}\,.
\label{eq:X1X2}
\end{eqnarray}
%%%%%%%%%%%%%%%%%%%%%%%%%%
%
Given $|\hat\gamma_u|/|\hat\beta_u|$, 
the value of the ratio of the constants 
$|\hat\alpha_u|/|\hat\beta_u|$ can be obtained using the 
value of the quark mass ratio $m_c/m_t$:
%%%%%%%%%%%%%%%%%%%%
\begin{equation}
\dfrac{m_c}{m_t} = |q_4|\,3\sqrt{2}\, 
\dfrac{|\hat\alpha_u|}{|\hat\beta_u|}\,
\left (1 + \frac{1}{8}\frac{|\hat\gamma_u|^2}
{|\hat\beta_u|^2}\right)^{-\,\frac{1}{2}} = 
10^{-2}\,A\, 3\sqrt{2}\, 
\dfrac{|\hat\alpha_u|}{|\hat\beta_u|}\,
\left (1 + \frac{1}{8}\frac{|\hat\gamma_u|^2}
{|\hat\beta_u|^2}\right)^{-\,\frac{1}{2}}\,. 
\label{eq:ab}
\end{equation}
%%%%%%%%%%%%%%%%%%%%%%
%

For the elements of the matrix $M_u^\dagger M_u\equiv M^{(2)}_u$, 
whose diagonalising matrix contributes to the CKM quark mixing matrix, 
we get keeping terms up to $\sim |q_4|^6 = \epsilon^6$:  
%%%%%%%%%%%%%%%%%%%%%%%%%%%%%%
\begin{eqnarray}
\nonumber
&& (M^{(2)}_u)_{11} \cong \left (\dfrac{v_u}{\sqrt{3}}\right )^2\,
36\, |\hat\alpha_u|^2\,|q_4|^4 + 256\,|q_4|^6\,
\left (2\,|\hat\beta_u|^2 +|\hat\gamma_u|^2\right) \,,\\[0.25cm]
\nonumber
&& (M^{(2)}_u)_{12}  \cong \left (\dfrac{v_u}{\sqrt{3}}\right )^2\,
\left[ -\,\frac{18}{\sqrt{2}}\,|\hat\alpha_u|^2\,|q_4|^2\,q^*_4
+ \dfrac{48}{\sqrt{2}}\,|q_4|^4q^*_4 
\left ( |\hat\gamma_u|^2 - 4\,|\hat\beta_u|^2\right)\right]\,,\\[0.25cm]
\nonumber
&& (M^{(2)}_u)_{13}  \cong \left (\dfrac{v_u}{\sqrt{3}}\right )^2\,
\left [
 2\sqrt{2}\,
(q^*_4)^3\,\left (-\,4\,|\hat\beta_u|^2 + |\hat\gamma_u|^2)\right)
 + \dfrac{72}{\sqrt{2}}\,|q_4|^4q_4\, |\hat\alpha_u|^2 
\right ]
,\\[0.25cm]
\nonumber
&& (M^{(2)}_u)_{22}  \cong \left (\dfrac{v_u}{\sqrt{3}}\right )^2\,
\frac{9}{2}\,|q_4|^2\left (|\hat\alpha_u|^2 + 8\,|\hat\beta_u|^2\,|q_4|^2 + 
|\hat\gamma_u|^2\,|q_4|^2\right )\,,\\[0.25cm]
\nonumber
&& (M^{(2)}_u)_{23}  \cong \left (\dfrac{v_u}{\sqrt{3}}\right )^2\,
\left [ 
3\,(q^*_4)^2\,\left ( -\,|\hat\beta_u|^2 + \frac{1}{8}\,|\hat\gamma_u|^2\right )
- 18\, q^*_4\,q_4^3\,|\hat\alpha_u|^2 \right ] \,,\\[0.25cm]
\nonumber
&& (M^{(2)}_u)_{33}  \cong \left (\dfrac{v_u}{\sqrt{3}}\right )^2\,
\left [
\frac{1}{4}\left (|\hat\beta_u|^2 + 
\frac{1}{8}\,|\hat\gamma_u|^2 \right ) + 
72\,|q_4|^6\,|\hat\alpha_u|^2 
\right ]\,,\\[0.25cm]  
&&(M^{(2)}_u)_{21} = (M^{(2)}_u)^*_{12}\,,\quad
(M^{(2)}_u)_{31} = (M^{(2)}_u)^*_{13}\,,\quad(M^{(2)}_u)_{32} = (M^{(2)}_u)^*_{23}\,,
\label{eq:MudegMu}
\end{eqnarray}
%%%%%%%%%%%%%%%%%%%%%%%%%%%%%%%%%%
%
where we have used the approximation $(1 + a |q_4|^4) \cong 1$ 
whenever the constant $|a| \leq 10^3$.

 It follows from the requirement that the up-quark mass 
ratios are correctly reproduced for $|q_4| = A\times 10^{-2}$ 
($\im [\tau] \sim 2$), with $A \sim~{a~few}$, 
that the ratios $|\hat\alpha_u|/|\hat\beta_u|$ and 
$|\hat\gamma_u|/|\hat\beta_u|$ are either much smaller than one 
or are much bigger than 1. For $\im[\tau] \sim 2.27391$, for example,
we have $A=2.8102$, and we get for the two possible ``solutions'' 
from Eqs. (\ref{eq:X1X2}) and (\ref{eq:ab}) using the best fit values of 
the ratios $m_u/m_t$ and $m_c/m_t$ given in Eqs. (\ref{eq:mumt}) 
and  (\ref{eq:mcmt}):
%%%%%%%%%%%%%%%%%%%%%%%%%%%%%%
\begin{eqnarray}
\label{eq:u1}
&& \left (\dfrac{|\hat\alpha_u|}{|\hat\beta_u|}\right)_1 \cong 19.2894\,,
\quad\quad\ \ \ 
\left (\dfrac{|\hat\gamma_u|}{|\hat\beta_u|}\right)_1 \cong 2266.4966\,,
\\[0.3cm]
\label{eq:u2}
&& \left (\dfrac{|\hat\alpha_u|}{|\hat\beta_u|}\right)_2 \cong 
2.407\times 10^{-2}\,,
\quad 
\left (\dfrac{|\hat\gamma_u|}{|\hat\beta_u|}\right)_2\cong 3.530\times 10^{-3}\,.
\end{eqnarray}
%%%%%%%%%%%%%%%%%%%%%%
%
From the statistical analyses we have performed it follows that 
only for the larger values of $|\hat\alpha_u|/|\hat\beta_u|$ and 
$|\hat\gamma_u|/|\hat\beta_u|$ 
the model might be viable: it definitely cannot describe correctly 
the quark mixing for the smaller values of the indicated constant ratios 
 and we do not discuss this possibility further.
 
It is not difficult to show that, e.g., 
for the ``benchmark'' values of the ratios of the constants
given in Eq. (\ref{eq:u1}), the terms 
i) $( -\,(18/\sqrt{2})\,|\hat\alpha_u|^2\,|q_4|^2\,q^*_4)$,
ii) $(72/\sqrt{2})|q_4|^4q_4\, |\hat\alpha_u|^2$, and 
iii) $18\, q^*_4\,q_4^3\,|\hat\alpha_u|^2$ in the expressions 
respectively of  $(M^{(2)}_u)_{12}$, $(M^{(2)}_u)_{13}$ and $(M^{(2)}_u)_{23}$ 
give insignificant contributions and can be neglected. 

Using the analytic expressions for the elements of the matrix 
$M^{(2)}_u$ in the case of ``large'' 
 $|\hat\alpha_u|/|\hat\beta_u|$ and 
$|\hat\gamma_u|/|\hat\beta_u|$ 
one can show that, in the 
approximation we are working 
with, $M^{(2)}_u$ can be cast in the form:
%%%%%%%%%%%%%%%%%%%%%%%%%%
\begin{equation}
M^{(2)}_u = P\,\tilde{M}^{(2)}_u\,P^\dagger\,, 
\label{eq:MuP}
\end{equation}
%%%%%%%%%%%%%%%%%%%%%%%%%%%%al 
%
where $\tilde{M}^{(2)}_u$ is a real symmetric matrix diagonalised by 
$\tilde{M}^{(2)}_u = O_u(M^{diag}_u)^2O^T_u$,  $O_u$ being 
an orthogonal matrix, $M^{diag}_u = \diag\,(m_u,m_c,m_t)$, 
and $P$ is a diagonal phase matrix:
%%%%%%%%%%%%%%%%%%%%%%%%%%
\begin{equation}
P = \diag\,(e^{-i\phi},1,e^{i2\phi})\,,~
{\rm with}~\phi = {\rm arg }[q_4] = \dfrac{\pi}{2}\,\re[\tau]\,,
\label{eq:P}
\end{equation}
%%%%%%%%%%%%%%%%%%%%%%%%%%%%%%
%
i.e., $\phi$ is the phase of $q_4$: $q_4 = |q_4|\,e^{i\phi}$.
The unitary matrix $U_u$, arising from the diagonalisation 
of the up-type quark mass matrix, which contributes to the CKM quark mixing 
matrix, $U_{\rm CKM} = U^\dagger_u\,U_d$, $U_d$ coming from the diagonalisation of 
the down-type quark mass term, is given thus by 
%%%%%%%%%%%%%%%%%%%%%%%%%%%%
\begin{equation}
U_u = P\,O_u\,.
\label{eq:Uu}
\end{equation}
%%%%%%%%%%%%%%%%%%

%%%%%%%%%%%%%%%%%%%%%%%%%%%%%%%
%
\subsection{Down-type Quarks}
%
%%%%%%%%%%%%%%%%%%%%%%%%%%%%%%%

As we have already noted, we assume that the RH down-type quark  
superfields $d^c$, $s^c$ and $b^c$ 
furnish the singlet representations 
$\mathbf{1}$, $\mathbf{1}$, $\mathbf{\hat{1}}$
of $S^\prime_4$ and have weights $k_{d^c}$, $k_{s^c}$,$k_{b^c}$. 
We assume further that $k_Q + k_{d^c} = 6$, 
$k_Q + k_{s^c} = 4$ and $k_Q + k_{b^c} = 7$.
Under these assumptions and recalling that the quark doublets 
are assumed to  form the triplet representation $\mathbf{3}$ 
of  $S^\prime_4$, it is possible to show that 
the down-type quark mass matrix is given by the following expression 
in the R-L convention:
%%%%%%%%%%%%%%%%%%%%%%%%%%%%%
\begin{equation}
    M_d =  \dfrac{v_d}{\sqrt{3}}
\begin{pmatrix}
\hat\alpha_d & 0  & 0 \\[2mm]
0 & \hat\beta_d & 0  \\[2mm]
0 & 0&  \hat\gamma_d  
\end{pmatrix}\,
\begin{pmatrix}
\left(Y^{(6)}_{\mathbf{3}}\right)_1 & \left(Y^{(6)}_{\mathbf{3}}\right)_3 & 
\left(Y^{(6)}_{\mathbf{3}}\right)_2
\\[2mm]
\left(Y^{(4)}_{\mathbf{3}}\right)_1 & \left(Y^{(4)}_{\mathbf{3}}\right)_3 & 
\left(Y^{(4)}_{\mathbf{3}}\right)_2
\\[2mm]
\left(\tilde{Y}^{(7)}_{\mathbf{\hat{3}'}}\right)_1 
& \left(\tilde{Y}^{(7)}_{\mathbf{\hat{3}'}}\right)_3 
& \left(\tilde{Y}^{(7)}_{\mathbf{\hat{3}'}}\right)_2
\end{pmatrix}\,,
\label{eq:MdY}
\end{equation}
%%%%%%%%%%%%%%%%%%%%%%%%%%%%%%%
%
where $v_d = v\cos\beta$ is the VEV of the SUSY Higgs field 
$H_d$ carrying a weak hypercharge (-1),
$Y^{(6)}_{\mathbf{3}}$, $Y^{(4)}_{\mathbf{3}}$ and
$\tilde{Y}^{(7)}_{\mathbf{\hat{3}'}}$ are weight 6, 4 and 7
level 4 modular forms furnishing 
the triplet representations 
$\mathbf{3}$, $\mathbf{3}$ and $\mathbf{\hat{3}'}$
of $S^\prime_4$. There are two triplet $\mathbf{\hat{3}'}$
modulat forms of weight 7. Thus,
%%%%%%%%%%%%%%%%%%%%%%%%%
\begin{equation}
\tilde{Y}^{(7)}_{\mathbf{\hat{3}'}} = 
g_d\, Y^{(7)}_{\mathbf{\hat{3}',1}} + Y^{(7)}_{\mathbf{\hat{3}',2}}\,,
\label{eq:Y7}
\end{equation}
%%%%%%%%%%%%%%%%%%%%%%%%%
%
where $g_d$ is a constant.
The constant parameters $\hat\alpha_d$, $\hat\beta_d$ and $\hat\gamma_d$
in Eq. (\ref{eq:MdY}) include the rescaling factors of the matter fields 
needed to cast their kinetic terms in canonical form: 
%%%%%%%%%%%%%%%%%%%%%
\begin{eqnarray}
\nonumber
 & \hat\alpha_d = 
\alpha_d\, (2\im [\tau])^3\,,\\[0.25cm]
\nonumber
& \hat\beta_d = \beta_d\,(2\im [\tau])^2\,,\\[0.25cm]
&\hat\gamma_d = \gamma_d\,(2\im [\tau])^{\frac{7}{2}}\,,
%
% \hat\alpha_{3,i} \rightarrow \alpha_{3,i}=
% \hat\alpha_{3,i}\, (2Im(\tau))^{\frac{5}{2}}\,,~i=1,2\,.
\label{eq:hatdconst}
\end{eqnarray}
%%%%%%%%%%%%%%%%%%%%%%%%%%%%%%
%
where $\alpha_d$, $\beta_d$ 
and $\gamma_d$ together with $g_d$ 
are the genuine constant parameters of the d-quark sector of 
the considered model. 

We do not impose the gCP-symmetry on the Lagrangian 
of the theory. The constants 
$\alpha_d$, $\beta_d$ and $\gamma_d$ can still be taken to be real 
without loss of generality: 
their phases can be absorbed by 
the fields $d^c$, $s^c$ and $b^c$.
However, the constant $g_d$ cannot be rendered real.
In the considered model $g_d$ plays the role of a complex 
parameter, which breaks the CP-symmetry explicitly.

For relatively large values of $\im [\tau]$, i.e., 
''close'' to the fixed point $\tau_\text{T} = i\infty$, one can show 
employing the method developed in  \cite{Novichkov:2021evw} 
that the down-type quark mass matrix $M_d$ has the following 
structure: 
%%%%%%%%%%%%%%%%%%%%%%%%%%%
\begin{equation}
M_{d} \sim \begin{pmatrix}
 \epsilon^2 & \epsilon & \epsilon^3 \\
 \epsilon^2& \epsilon & \epsilon^3 \\
 \epsilon^3 & \epsilon^2 & 1
\end{pmatrix}\,.
\end{equation}
%%%%%%%%%%%%%%%%%%%%%%
%
It leads to a hierarchical 
down-type quark mass spectrum 
{\bf $(m_b:m_s:m_d) \sim m_b (1:\epsilon:\epsilon^2)$.}

Using the $q$-expansions of the modular forms 
$Y^{(6)}_{\mathbf{3}}$, $Y^{(4)}_{\mathbf{3}}$,
$Y^{(7)}_{\mathbf{\hat{3}',1}}$ and $Y^{(7)}_{\mathbf{\hat{3}',2}}$ 
present in $M_d$, it is possible to derived 
analytic expressions for the d-quark 
mass matrix. Employing again the 
normalisation of the relevant modular
forms introduced in\cite{Novichkov:2020eep}   
and keeping terms not smaller than $|q_4|^4$
we find in 
the R-L convention:
%%%%%%%%%%%%%%%%%%%%%%%%%%%%%
\begin{equation}
    M_d =  \dfrac{v_d}{\sqrt{3}}\dfrac{3}{\sqrt{2}}
\begin{pmatrix}
\hat\alpha_d & 0  & 0 \\[2mm]
0 & \hat\beta_d & 0  \\[2mm]
0 & 0&  \hat\gamma_d  
\end{pmatrix}\,
\begin{pmatrix}
 2\sqrt{2}\,q^2_4 & \dfrac{1}{2}\,q_4 & 10\,q^3_4 \\
 2\sqrt{2}\,q^2_4 & -\,q_4 & 4\,q^3_4 \\
 2\sqrt{2}\,q^3_4 \left (\dfrac{7}{\sqrt{37}}g_d + 3\right ) & -\,3\,q^2_4 
& -\,\dfrac{g_d}{4\sqrt{37}}(1+252\,q^4_4) + 12\,q^4_4
\end{pmatrix}\,.
\label{eq:Mdq4}
\end{equation}
%%%%%%%%%%%%%%%%%%%%%%
%
For the masses of the 
$d$-, $s$- and $b$-quarks using the expression for 
$M_d$ we get: 
%%%%%%%%%%%%%%%%%%%%%%%%%%%
\begin{eqnarray}
\nonumber
&& m_d = \dfrac{v_d}{\sqrt{3}}\,9\,|q_4|^2\,
\dfrac{|\hat\alpha_d\,\hat\beta_d|}
{
\sqrt{
\dfrac{|\hat\alpha_d|^2}{4} + |\hat\beta_d|^2
}
}\,,\\[0.25cm]
\nonumber
&& m_s = \dfrac{v_d}{\sqrt{3}}\,\dfrac{3}{\sqrt{2}}\,
|q_4|\,
\sqrt{
\dfrac{|\hat\alpha_d|^2}{4} + |\hat\beta_d|^2
}\,,\\[0.25cm]
&& m_b = \dfrac{v_d}{\sqrt{3}}\,\dfrac{3}{\sqrt{2}}\, 
\dfrac{|g_d\,\hat\gamma_d|}{4\sqrt{37}}\,.
\label{eq:mdmsmb}
\end{eqnarray}
%%%%%%%%%%%%%%%%%%%%%%%
%

Analogously to the up-type quark mass ratios, 
for the ratios $m_d/m_b$ and $m_s/m_b$ 
we will use their values 
at the GUT scale $2\times 10^{16}$ GeV 
obtained taking into account the RG running effects 
for $\tan\beta = 10$ \cite{Antusch:2025fpm}: 
%%%%%%%%%%%%%%%%%%%%%%%%%5
\begin{eqnarray}
\label{eq:mdmb}
& \dfrac{m_d}{m_b} = (8.99 \pm 0.277)\times 10^{-4}\,,\\[0.25cm]
\label{eq:msmb}
& \dfrac{m_s}{m_b} = (1.80  \pm 0.037 ) \times 10^{-2}\,.
\end{eqnarray}
%%%%%%%%%%%%%%%%%%%%%
%

The value of the ratio of the constants 
$|\hat\alpha_d|$ and $|\hat\beta_d|$
can be determined using the value of
the quark mass ratio $m_d/m_s$. 
From Eqs. (\ref{eq:mdmb}) and 
(\ref{eq:msmb}) we get 
$m_d/m_s = F_{ds}\times 10^{-2}$, 
with  $F_{ds} = 4.9944$. 
If we denote by
$Z_d =|\hat\beta_d|/|\hat\alpha_d|$ and 
set $|q_4| = A\,\times 10^{-2}$,
the ratio $Z_d$ satisfies the following quadratic equation:
$12\sqrt{2}\,A\,Z_d = 4\,F_{ds}\,Z_d^2 + F_{ds}$.
The two solutions $Z_{d(1,2)}$ are given approximately by:
%%%%%%%%%%%%%%%%%%%%%%%%%5
\begin{equation}
 Z_{d(1,2)} = \dfrac{3\sqrt{2}A \mp \sqrt{18\,A^2 - F^2_{ds}}}
{2F_{ds}}\,.
\label{eq:Z1Z2}
\end{equation}
%%%%%%%%%%%%%%%%%%%%%%%%%%
%
The solutions exist provided $18\,A^2 \geq  F^2_{ds}$ or
$A = |q_4|\times 10^2 \geq F_{ds}/(3\sqrt{2}) \cong 1.1909$. %=5.0526/4.24261$.

Given $|q_4|$ and the ratio $|\hat\beta_d|/|\hat\alpha_d|$, 
the value of the ratio of the constants 
$|\hat\alpha_d|/|g_d\hat\gamma_d|$ 
can be derived using the value of the ratio of the strange and 
bottom quark masses, $m_s/m_b$:
%%%%%%%%%%%%%%%%%%%%%%%%%%
\begin{equation}
\dfrac{|\hat\alpha_d|}{|g_d\hat\gamma_d|} = 
\dfrac{m_s}{m_b}\,
\left (
|q_4|\,2\,\sqrt{37}\,
\sqrt{
1 + \dfrac{4|\hat\beta_d|^2}{|\hat\alpha_d|^2}
}
\right)^{-\,1}\,.
\label{eq:adgammad}
\end{equation} 
%%%%%%%%%%%%%%%%%%%%%%%%%%%%%%%%
%

For the ratios $|\hat\beta_d|/|\hat\alpha_d|$
and $|\hat\alpha_d|/|g_d\hat\gamma_d|$ we get 
using $F_{ds} = 4.9944$ and $A=2.8102$ and the best fit values of 
$m_d/m_s$ and $m_s/m_b$:
%%%%%%%%%%%%%%%%%%%%%%%%%%%%%%
\begin{eqnarray}
\label{eq:d1}
&& \left (\dfrac{|\hat\beta_d|}{|\hat\alpha_d|}\right)_1 \cong 0.109774\,,
\quad 
\left (\dfrac{|\hat\alpha_d|}{|g_d\hat\gamma_d|}\right)_1 \cong 0.05142\,,
\\[0.3cm]
\label{eq:d2}
&& \left (
\dfrac{|\hat\beta_d|}{|\hat\alpha_d|}
\right)_2 \cong 
2.2774\,,
\qquad 
\left (
\dfrac{|\hat\alpha_d|}{|g_d\hat\gamma_d|}
\right)_2 \cong 0.01129\,.
\end{eqnarray}
%%%%%%%%%%%%%%%%%%%%%%
%
From the statistical analysis of the model we have performed 
it follows that only for the values of 
$|\hat\beta_d|/|\hat\alpha_d|$
and $|\hat\alpha_d|/|g_d\hat\gamma_d|$ given in Eq.(\ref{eq:d1}) 
it might be possible to describe the data on the quark mixing 
and CP-violation and we are not going to consider the alternative values 
quoted in Eq. (\ref{eq:d2}) further.

For the elements of the matrix $M_d^\dagger M_d\equiv M^{(2)}_d$, 
whose diagonalising matrix contributes to the CKM quark mixing matrix, 
we get keeping terms up to $\sim |q_4|^4 = \epsilon^4$:  
%%%%%%%%%%%%%%%%%%%%%%%%%%%%%%
\begin{eqnarray}
\nonumber
&& (M^{(2)}_d)_{11} \cong \left (\dfrac{v_d}{\sqrt{3}}\right )^2\,
\left (\dfrac{3}{\sqrt{2}} \right)^2
\left[ \,8\,|q_4|^4\, 
\left(|\hat\alpha_d|^2 + |\hat\beta_d|^2\right ) 
+ |q_4|^6\,|\hat\gamma_d|^2\,
\left |\dfrac{56\sqrt{2}}{4\sqrt{37}}\,g_d 
+ 6\sqrt{2}
\right |^2 
\right]\,,\\[0.25cm]
\nonumber
&& (M^{(2)}_d)_{12}  \cong \left (\dfrac{v_d}{\sqrt{3}}\right )^2\,
\left (\dfrac{3}{\sqrt{2}} \right)^2\,q_4\,(q^2_4)^*\,
\sqrt{2}\, \left[ |\hat\alpha_d|^2 - 2\,|\hat\beta_d|^2
-\,|q_4|^2\,6\,\dfrac{|g_d\hat\gamma_d|^2}{g_d}\left(\dfrac{7}{\sqrt{37}} 
+ \dfrac{3}{g^*_d}\right)  \right]\,,\\[0.25cm]
\nonumber
&& (M^{(2)}_d)_{13}  \cong -\,\left (\dfrac{v_d}{\sqrt{3}}\right )^2\,
\left (\dfrac{3}{\sqrt{2}}\right)^2\,(q^3_4)^*\,
\dfrac{\sqrt{2}\,g_d\,|\hat\gamma_d|^2}{4\sqrt{37}}\,
\left [ \dfrac{14\,g^*_d}{\sqrt{37}} + 6\right ]\,,\\[0.25cm]
\nonumber
&& (M^{(2)}_d)_{22}  \cong \left (\dfrac{v_d}{\sqrt{3}}\right )^2\,
\left (\dfrac{3}{\sqrt{2}}\right)^2\,
|q_4|^2\left (\dfrac{|\hat\alpha_d|^2}{4} + |\hat\beta_d|^2
 + 9\,|q_4|^2\,|\hat\gamma_d|^2\right )\,,\\[0.25cm]
\nonumber
&& (M^{(2)}_d)_{23}  \cong \left (\dfrac{v_d}{\sqrt{3}}\right )^2\,
\left (\dfrac{3}{\sqrt{2}}\right)^2\,
\left (
(q^2_4)^*\,\dfrac{3\,g_d}{4\sqrt{37}}\,|\hat\gamma_d|^2 
+ q^*_4\,q^3_4\,\left [ 5|\hat\alpha_d|^2 - 4\,|\hat\beta_d|^2\right]
\right )
\,,\\[0.25cm]
\nonumber
&& (M^{(2)}_d)_{33}  \cong \left (\dfrac{v_d}{\sqrt{3}}\right )^2\,
\left (\dfrac{3}{\sqrt{2}}\right)^2\,
|\hat\gamma_d|^2 \left | \dfrac{g_d}{4\sqrt{37}}(1 + 9\times 28\,q^4_4) 
 -\,12\,q^4_4 \right |^2\,,\\[0.25cm]  
&& (M^{(2)}_d)_{21} = (M^{(2)}_d)^*_{12}\,,\quad
(M^{(2)}_d)_{31} = (M^{(2)}_d)^*_{13}\,,\quad (M^{(2)}_d)_{32} = (M^{(2)}_d)^*_{23}\,.
\label{eq:MddegMd}
\end{eqnarray}
%%%%%%%%%%%%%%%%%%%%%%%%%%%%%%%%%%
%
We kept the term $\sim |q_4|^6$ in $(M^{(2)}_d)_{11}$ because it is 
multiplied by a factor which is much larger than the factor multiplying the 
term  $\sim |q_4|^4$.

%%%%%%%%%%%%%%%%%%%%%%%%%%%%%%%
%
\subsection{CP-Symmtry Violation}
%
%%%%%%%%%%%%%%%%%%%%%%%%%%%%%%%

As we will see in what follows, the description of the d-quark mass ratios 
and of, e.g., the quark mixing angle $\theta_{12}$ - the Cabbibo angle - 
is only possible for $|q_4| \ll 1$,  
$|\hat\beta_d|/|\hat\alpha_d| \ll 1$ and 
$|\hat\alpha_d|/|g_d\hat\gamma_d| \ll 1$.
As a consequence, the term $\propto q^*_4\,q^3_4$ in the expression 
for $(M^{(2)}_d)_{23}$ is much smaller than the term 
$\propto (q^2_4)^*$ and can be neglected. 
It follows from Eq. (\ref{eq:MddegMd}) that 
under the indicated conditions and using  $q_4 = |q_4|e^{i\phi}$ 
the matrix  $M^{(2)}_d$ can be cast in the form:
%%%%%%%%%%%%%%%%%%%%%%%%%%%%%%%%
\begin{equation}
M^{(2)}_d = 
P\,\tilde{M}^{(2)}_d\,P^*\,,
\label{eq:tM2d}
\end{equation}
%%%%%%%%%%%%%%%%%%%%%%%%%%%%%%%%
%
where the diagonal phase matrix $P$  is defined in Eq. (\ref{eq:P}), and 
%%%%%%%%%%%%%%%%%%%%%%%
\begin{eqnarray}
\tilde{M}^{(2)}_d = 
\left( \dfrac{v_d}{\sqrt{3}}\right )^2\,
\left (\dfrac{3}{\sqrt{2}} \right)^2\,
\begin{pmatrix}
a_{11} & a_{12} & a_{13} \\
a^*_{12} & a_{22} & a_{23} \\
a^*_{13} & a^*_{23}  & a_{33}
\end{pmatrix}\,. 
\label{eq:M2dP}
\end{eqnarray}
%%%%%%%%%%%%%%%%%%%%%%
%
The elements $a_{11}$, $a_{22}$ and $a_{33}$ 
of the matrix $\tilde{M}^{(2)}_d$ are real, 
while $a_{12}$,  $a_{13}$ and $a_{23}$ are complex as a 
consequence of the constant $g_d$ being complex.
 Thus,  $\tilde{M}^{(2)}_d = (\tilde{M}^{(2)}_d)^\dagger$ and
is diagonalised by a unitary matrix $\tilde{U}_d$:
 $\tilde{M}^{(2)}_d = \tilde{U}_d\, (M^{diag}_d)^2\,(\tilde{U}_d)^\dagger$,
where $M^{diag}_d = \diag\,(m_d,m_s,m_b)$. 
Correspondingly, the unitary matix $U_d$ which contributes 
to the CKM matrix is given by: 
%%%%%%%%%%%%%%%%%%%%%%%%%
\begin{equation}
U_d = P\,\tilde{U}_d\,,~~\tilde{U}_d-{\rm unitary}\,.
\label{eq:Ud0}
\end{equation}
%%%%%%%%%%%%%%%%%%%%%%%%

From Eqs. (\ref{eq:Uu}) and  (\ref{eq:Ud0}) we get for the CKM matrix:
%%%%%%%%%%%%%%%%%
\begin{equation}
U_{\rm CKM} = U^\dagger_u\,U_d =  O^T_u\,P^*\,P\,\tilde{U}_d
=  O^T_u\,\tilde{U}_d\,.
\label{eq:UCKM}
\end{equation}
%%%%%%%%%%%%%%%%%%%%%%%%
%
It follows from this result that in the approximation we have adopted 
keeping terms not smaller than $|q_4|^4$ in the up- and down-type quark mass 
matrices, the CKM quark mixing matrix does not depend on the real part 
of the modulus $\tau$. This in turns implies that the  CP-violation in 
the quark sector in the model we are considering is practically 
entirely due to the explicit breaking of the CP symmetry by the complex 
value of the constant $g_d$. Although the VEV of $\tau$ can also violate the CP 
symmetry, its contribution to the 
{bf observed} non-conservation of the 
CP-symmetry in the quark sector is essentially negligible 
\footnote{We have already encountered in \cite{Petcov:2022fjf} 
a similar situation of mutual cancelation of the CP-violating contributions 
from the up-quark and down-quark sectors to the to CKM matrix, originating 
from the CP-violating VEV of the modulus $\tau$.
}.
In view of this result in the analyses which follow we set 
$\re[\tau] = 0$
\footnote{If the considered quark flavour model is embedded 
in a larger modular model of both quark and lepton flavours,
$\re[\tau]$ can be used as an additional parameter in the description 
of lepton masses and mixing and can be a source of CP-violation 
in the lepton sector.
}. In this case the only complex elements in the matrix 
$M^{(2)}_d$ are 
$(M^{(2)}_d)_{12} = |(M^{(2)}_d)_{12}|e^{i\phi_{12}}$, 
$(M^{(2)}_d)_{13} = |(M^{(2)}_d)_{13}|e^{i\phi_{13}}$, 
$(M^{(2)}_d)_{23} = |(M^{(2)}_d)_{23}|e^{i\phi_{23}}$,
 $(M^{(2)}_d)_{21} = ((M^{(2)}_d)_{12})^*$,
$(M^{(2)}_d)_{31} = ((M^{(2)}_d)_{13})^*$ and
 $(M^{(2)}_d)_{32} = ((M^{(2)}_d)_{23})^*$, 
where
%%%%%%%%%%%%%%%%%%%%%%%%%%%%%%%%
\begin{equation}
\phi_{12} =  \arg[1 - y(g_d)]\,,\quad
\phi_{13} = \arg\left[ \dfrac{7}{\sqrt{37}} + \dfrac{3}{g^*_d}\right]\,,\quad
 \phi_{23} = \arg [g_d]\,.
\label{eq:ph13p23} 
\end{equation}
%%%%%%%%%%%%%%%%%%%%%%%%%%%%%
%
with 
%%%%%%%%%%%%%%%%%%%%%%%%%%
\begin{equation}
y(g_d) = |q_4|^2\,6\,\dfrac{|g_d\hat\gamma_d|^2}{|\hat\alpha_d|^2}
\dfrac{1}{g_d}\left(\dfrac{7}{\sqrt{37}} 
+ \dfrac{3}{g^*_d}\right)\,.
\label{eq:ygd}
\end{equation}
%%%%%%%%%%%%%%%%%%%%%%%%%%%%%%%
%
\section{Mixing from the Down-type Quarks}
\label{sec:dmixing}
%
%%%%%%%%%%%%%%%%%%%%%%%%%%%%%%%
%

Under the conditions specified in the preceding subsection 
(keeping terms not smaller than $|q_4|^4$, setting 
$\re [\tau] = 0$, $|q_4| \ll 1$,  
$|\hat\beta_d|/|\hat\alpha_d| \ll 1$ and 
$|\hat\alpha_d|/|g_d\hat\gamma_d| \ll 1$), it is possible to diagonalise 
the matrix $M^{(2)}_d$ and obtain analytic expression for the 
unitary matrix $\tilde{U}_d = U_d$ contributing to the CKM matrix 
Eq. (\ref{eq:Ud0}), with $P$ being now the unit matrix
 since $\re[\tau] = 0$.
We get:
%%%%%%%%%%%%%%%%%%%%%%%%%%%%%%%%
\begin{eqnarray}
\nonumber
& M^{(2)}_d =  U_d\,M^{(2,diag)}_d\,U^\dagger\,,\\[0.25cm]
& U_d = P_{23}\,O_{23}\,P_{13}\,O_{13}\,P_{12}\,O_{12}\,,
\label{eq:M2ddiag}
\end{eqnarray}
%%%%%%%%%%%%%%%%%%
%
where $M^{(2,diag)}_d = \diag\,(m^2_d,m^2_s,m^2_b)$,
$P_{23} = \diag\,(1,1,e^{-\,i\phi_{23}})$,
$P_{13} = \diag\,(1,1,e^{-\,i(\phi_{13}-\phi_{23})})$
and $P_{12} = \diag\,(1,e^{-\,i \tilde{\phi}_{12}},1)$ are diagonal phase 
matrices, $\tilde{\phi}_{12}$ is defined below, 
and $O_{23}$, $O_{13}$ and $O_{12}$ are orthogonal matrices 
describing 2-3, 1-3 and 1-2 rotations: 
%%%%%%%%%%%%%%%%%%%%%%%%%
\begin{eqnarray}
O_{23} = \begin{pmatrix}
1 & 0 & 0 \\
0 & c^d_{23} & s^d_{23} \\
0 &-\,s^d_{23}  & c^d_{23}
\end{pmatrix}\,,\quad
O_{13}= \begin{pmatrix}
c^d_{13} & 0 & s^d_{13} \\
0 & 1 & 0 \\
-\,s^d_{13} & 0  & c^d_{13}
\end{pmatrix}\,,\quad
O_{12}= \begin{pmatrix}
c^d_{12} & s^d_{12} & 0 \\
-\,s^d_{12} & c^d_{12} & 0 \\
0 & 0  & 1
\end{pmatrix}\,. 
\label{eq:Oij}
\end{eqnarray}
%%%%%%%%%%%%%%%%%%%%%%%%%%
%
Here $c^d_{ij} = \cos\theta^d_{ij}$ 
and $s^d_{ij} = \sin\theta^d_{ij}$ 
with $ij = 23,13,12$. 
Given the hierarchy of the elements of 
$M^{(2)}_d$, the angles are determined by:
%%%%%%%%%%%%%%%%%%%%%%%%%%
\begin{eqnarray}
\nonumber
&&\tan2\theta^d_{23} = 
\dfrac{2|(M^{(2)}_d)_{23}|}{(M^{(2)}_d)_{33} - (M^{(2)}_d)_{22}}\,,\\[0.3cm]
\nonumber
&&\tan2\theta^d_{13} = -\,
\dfrac{2|(M^{(2)}_d)_{13}|}{(M^{(2)}_d)_{33} - (M^{(2)}_d)_{11}}\,,\\[0.3cm]
&&\tan2\theta^d_{12} = 
\dfrac{2|(\tilde{M}^{(2)}_d)_{12}|}{(\tilde{M}^{(2)}_d)_{22} - 
(\tilde{M}^{(2)}_d)_{11}}\,,
\label{eq:tan2thdij} 
\end{eqnarray}
%%%%%%%%%%%%%%%%%%%%%%%%%
%
where $(\tilde{M}^{(2)}_d)_{12} = |(\tilde{M}^{(2)}_d)_{12}|e^{i\tilde{\phi}_{12}}$,
$(\tilde{M}^{(2)}_d)_{22}$ and 
$(\tilde{M}^{(2)}_d)_{11}$ include corrections in the $(M^{(2)}_d)_{12}$, 
$(M^{(2)}_d)_{22}$ and 
$(M^{(2)}_d)_{11}$ elements due to the 2-3 and 1-3 rotations, respectively: 
%%%%%%%%%%%%%%%%%%%%%%%%%%%
\begin{eqnarray}
\nonumber
&& (\tilde{M}^{(2)}_d)_{12} = (M^{(2)}_d)_{12}\cos\theta^d_{23} - 
(M^{(2)}_d)_{13}e^{-\,i\phi_{23}}\sin\theta^d_{23} 
\cong 
(M^{(2)}_d)_{12} - (M^{(2)}_d)_{13}e^{-\i\phi_{23}}\,\sin\theta^d_{23}\,,
\\[0.3cm]
\nonumber
&& (\tilde{M}^{(2)}_d)_{22} = (M^{(2)}_d)_{22} - \dfrac{|(M^{(2)}_d)_{23}|^2}
{ (M^{(2)}_d)_{33} - (M^{(2)}_d)_{22} }  \,,\\[0.3cm]
&& (\tilde{M}^{(2)}_d)_{11} = (M^{(2)}_d)_{11}
 - \dfrac{|(M^{(2)}_d)_{13}|^2}
{(M^{(2)}_d)_{33}} \,.
\label{eq:tMd22tMd11}
\end{eqnarray}
%%%%%%%%%%%%%%%%%%%%%%%%%%
%
The corrections in all three elements   $(M^{(2)}_d)_{12}$,
$(M^{(2)}_d)_{22}$ and $(M^{(2)}_d)_{11}$ are  important. 
It follows from the expression for $(\tilde{M}^{(2)}_d)_{12}$ that
%%%%%%%%%%%%%%%%%%%%%%%%%%%%
\begin{equation}
\tilde{\phi}_{12} \equiv
\arg[(\tilde{M}^{(2)}_d)_{12}] = 
\arg\left[{(M^{(2)}_d)_{12}} -
(M^{(2)}_d)_{13}e^{-\,i\phi_{23}}\,\sin\theta^d_{23} \right]\,.
\label{eq:tphi12}
\end{equation}
%%%%%%%%%%%%%%%%%%%%%%%%%%
%
Using the expressions for $(M^{(2)}_d)_{13}$ and $(M^{(2)}_d)_{12}$
given in Eq. (\ref{eq:MddegMd}) 
% and neglecting the term 
% $|\hat\beta_d|^2/|\hat\alpha_d|^2 \ll 1$ in  $(M^{(2)}_d)_{12}$,
we can cast $(\tilde{M}^{(2)}_d)_{12}$ in the form:
%%%%%%%%%%%%%%%%%%%%%%%%%%%%%%%%%
\begin{eqnarray}
&&(\tilde{M}^{(2)}_d)_{12}
\cong \left(
\dfrac{v_u}{\sqrt{3}}
\right )^2\,
\left(
\dfrac{3}{\sqrt{2}} \right)^2 
\sqrt{2}\,|q_4|^3\,|\hat\alpha_d|^2\,\nonumber\\
&&\hskip 1.5 cm \times \left ( 1 -\,2\,\dfrac{|\hat\beta_d|^2}{|\hat\alpha_d|^2} 
-\,\dfrac{|g_d\hat\gamma_d|^2}{|\hat\alpha_d|^2}\,
\left (\dfrac{7}{\sqrt{37}} + \dfrac{3}{g^*_d}\right) 
\left (6\,\dfrac{|q_4|^2}{g_d} 
- \dfrac{\sin\theta^d_{23}e^{-\,i\phi_{23}}}{2\sqrt{37}}\right)
\right )\,.
\label{eq:tM12d1}
\end{eqnarray}
%%%%%%%%%%%%%%%%%%%%%%%%%%%%%%%
%

For $\sin\theta^d_{23}$ we get from Eqs. (\ref{eq:MddegMd}), 
(\ref{eq:tan2thdij}) and (\ref{eq:tMd22tMd11}):
%%%%%%%%%%%%%%%%%%%%%%%%%%%%%%%%%%%%
\begin{equation}
\sin\theta^d_{23} = 
\dfrac{1}{\sqrt{2}} 
\left ( 1 - \dfrac{1}
{
\sqrt{1 + |q_4|^4\,(4\sqrt{37})^2 \dfrac{36}{|g_d|^2}}
}
\right)^{\frac{1}{2}} 
\cong 12\,\sqrt{37}\,\dfrac{|q_4|^2}{|g_d|}\,.
\label{eq:sth23d}
\end{equation}
%%%%%%%%%%%%%%%%%%%%%
%
In the case of  $\sin\theta^d_{13}$ we obtain:
%%%%%%%%%%%%%%%%%%%%%%%%%%%%%%%%%%%%
\begin{equation}
\sin\theta^d_{13} = 
-\,\dfrac{1}{\sqrt{2}}\, 
\left ( 1 - 
\dfrac{1}
{
\sqrt{
1 + |q_4|^6\,32\,(4\sqrt{37})^2 
\left | \dfrac{7}{\sqrt{37}} + \dfrac{3e^{i\phi_{23}}}{|g_g|} \right |^2
}
}
\right)^{\frac{1}{2}} 
\cong -\,|q_4|^3
8\,\sqrt{74}\,
\left |\dfrac{7}{\sqrt{37}} + \dfrac{3e^{i\phi_{23}}}{|g_g|}\right |\,.
\label{eq:sth13d}
\end{equation}
%%%%%%%%%%%%%%%%%%%%%
%
Inserting the result for  $\sin\theta^d_{23}$ 
and Eq. (\ref{eq:tM12d1}) simplifies the expression for 
$(\tilde{M}^{(2)}_d)_{12}$ to:
%%%%%%%%%%%%%%%%%%%%%%%%%%%%%%%%%
\begin{equation}
(\tilde{M}^{(2)}_d)_{12}
\cong \left(\dfrac{v_u}{\sqrt{3}}\right )^2\,
\left(\dfrac{3}{\sqrt{2}} \right)^2 
\sqrt{2}\,|q_4|^3\,|\hat\alpha_d|^2\,
\left( 1 - y^\prime(g_d)  -\, 2\,\dfrac{|\hat\beta_d|^2}{|\hat\alpha_d|^2}  
\right)\,,
\label{eq:tM12d2}
\end{equation}
%%%%%%%%%%%%%%%%%%%%%%%%%%
%
where
%%%%%%%%%%%%%%%%%%%%%%%%%%
\begin{equation}
 y^\prime(g_d) = 
  6\,|q_4|^2\,\dfrac{|g_d\hat\gamma_d|^2}{|\hat\alpha_d|^2}\,
\left (\dfrac{7}{\sqrt{37}} + \dfrac{3}{g^*_d}\right) 
\left (\dfrac{1}{g_d} - \dfrac{e^{-\,i\phi_{23}}}{|g_d|}\right) = 0\,,
\label{eq:yd}
\end{equation}
%%%%%%%%%%%%%%%%%%%%%%%%%%%%%%%
%
where we have used the fact that $\phi_{23} = \arg[g_d]$. 
Thus, $((\tilde{M}^{(2)}_d)_{12})^* = (\tilde{M}^{(2)}_d)_{12}$.

The expressions for  $(\tilde{M}^{(2)}_d)_{22}$ and 
$(\tilde{M}^{(2)}_d)_{11}$ also simplify considerably due to the 
corrections from the 2-3 and 1-3 rotations, respectively.
Using the results given  in Eqs.(\ref{eq:MddegMd})
and (\ref{eq:tMd22tMd11}) we obtain:
%%%%%%%%%%%%%%%%%%%%%%%%%%%%%%%%%%%
\begin{eqnarray}
\nonumber
&& (\tilde{M}^{(2)}_d)_{22} 
\cong \dfrac{
(M^{(2)}_d)_{22}
\left ( 1 +  4\,\dfrac{|\hat\beta_d|^2}{|\hat\alpha_d|^2}\right )
}
{1 + X_{22}  + 4\, \dfrac{|\hat\beta_d|^2}{|\hat\alpha_d|^2}
} 
 = \left(\dfrac{v_u}{\sqrt{3}}\right )^2\,
\left(\dfrac{3}{\sqrt{2}} \right)^2\,
 |q_4|^2\,\dfrac{|\hat\alpha_d|^2}{4} 
\left ( 1 + 4\, \dfrac{|\hat\beta_d|^2}{|\hat\alpha_d|^2} \right )\,,\\[0.25cm]
\nonumber
&& (\tilde{M}^{(2)}_d)_{11} 
 = (M^{(2)}_d)_{11}\left (1 - \dfrac{|(M^{(2)}_d)_{13}|^2}
 {(M^{(2)}_d)_{11}(M^{(2)}_d)_{33}} \right) \,
\cong \dfrac{(M^{(2)}_d)_{11}
\left ( 1 +  \dfrac{|\hat\beta_d|^2}{|\hat\alpha_d|^2}\right) }
{1 + X_{11} +  \dfrac{|\hat\beta_d|^2}{|\hat\alpha_d|^2}   
}\,\\[0.25cm] 
&& \hskip 1.5 cm \cong \left(\dfrac{v_u}{\sqrt{3}}\right )^2\,
\left(\dfrac{3}{\sqrt{2}} \right)^2\,
8\,|q_4|^4\,|\hat\alpha_d|^2 
\left( 1 + \dfrac{|\hat\beta_d|^2}{|\hat\alpha_d|^2}\right)\,,
\label{eq:tM22tM11}
\end{eqnarray}
%%%%%%%%%%%%%%%%%%%%%%%%
%
where 
%%%%%%%%%%%%%%%%%%%%%%%%%%%%%%
\begin{eqnarray}
\nonumber
&&  X_{22} = |q_4|^2\,36\,\dfrac{|\hat\gamma_d|^2}{|\hat\alpha_d|^2}\,,
\\[0.25cm]
&& X_{11} = |q_4|^2\dfrac{|g_d\hat\gamma_d|^2}{|\hat\alpha_d|^2}\,
\left |\dfrac{7}{\sqrt{37}} + \dfrac{3}{g^*_g}\right |^2\,,
\label{eq:XdYd}
\end{eqnarray}
%%%%%%%%%%%%%%%%%%%%
%
and we have used the fact in
Eq.(\ref{eq:MddegMd}), 
$(M^{(2)}_d)_{22} \cong C\,|q_4|^2\,|\hat\alpha_d|^2\,
(1 + X_{22} + 4|\hat\gamma_d|^2/|\hat\alpha_d|^2)/4$ 
and $(M^{(2)}_d)_{11} \cong 
C\,|q_4|^4\,8\,|\hat\alpha_d|^2\,(1 + X_{11} + 
|\hat\beta_d|^2/|\hat\alpha_d|^2)$
with $C\equiv (v_d/\sqrt{3})^2\,(3/\sqrt{2})^2$.

Finally, for $\tan2\theta^d_{12}$ and $\sin\theta^d_{12}$ 
we get from Eqs. (\ref{eq:MddegMd}), 
(\ref{eq:tan2thdij}), (\ref{eq:tMd22tMd11}), (\ref{eq:tM12d2}) 
(\ref{eq:yd}) and (\ref{eq:tM22tM11}): 
%%%%%%%%%%%%%%%%%%%%%%%%%%%%%%%%%%%%
\begin{equation}
\tan2\theta^d_{12} = 2 \dfrac{
4\sqrt{2}\,|q_4|\,\left(1 - 2r^2_d)\right) 
}
{1 - 32\,|q_4|^2 + 4r^2_d}\,,~~
%
% \tan2\theta^d_{12} = 2 \dfrac{
% 4\sqrt{2}\,|q_4|\,
% \left(1 - 2 \dfrac{|\hat\beta_d|^2}{|\hat\alpha_d|^2}\right)
% }
% {1 - 32\,|q_4|^2 + 4\,\dfrac{|\hat\beta_d|^2}{|\hat\alpha_d|^2}}\,,~~
% \quad 
% \sin\theta^d_{12} = \dfrac{4\sqrt{2}\,|q_4|}{\sqrt{1 + 32\,|q_4|^2}} 
%\left(1 - 12 \dfrac{|\hat\beta_d|^2}{|\hat\alpha_d|^2}\right)^{\frac{1}{2}}
% + O(10^{-4})\,.
\sin\theta^d_{12} = 
\dfrac{4\sqrt{2}\,|q_4|\left (1-8r^2_d+36r^4_d\right)^{\frac{1}{2}} }
{\left(1+32\,|q_4|^2(1-16r^2_d+72r^4_d)+4r^2_d\right)^{\frac{1}{2}} } + O(10^{-4})\,,
\label{eq:tan2th12d}
\end{equation}
%%%%%%%%%%%%%%%%%%%%%
% 
where $r_d \equiv |\hat\beta_d|/|\hat\alpha_d|$ and we have used 
$\sin\theta^d_{12} = (1-\cos2\theta^d_{12})^{\frac{1}{2}}/\sqrt{2}$.
% \\
% {\bf Serguey: the difference between the value 
% of $\sin\theta^d_{12}$ calculated using the expression 
% for $\cos2\theta^d_{12}$ obtained from the 
% one for $\tan2\theta^d_{12}$ without approximations 
% and the approximate expression for $\sin\theta^d_{12}$
% given the above equation is $1.31\times 10^{-4}$. 
% By adding one more known term, present in both the numerator and denominator 
% of the approximate expression for $\sin\theta^d_{12}$, 
% which I have neglected,  this difference can be reduced 
% to $7.901\times 10^{-5}$, 
% but the approximate expression for $\sin\theta^d_{12}$ becomes 
% somewhat more cumbersome.}\\

Equations  (\ref{eq:tphi12}), 
(\ref{eq:tM12d2}) and (\ref{eq:yd})
imply also that $\tilde{\phi}_{12} = 0$. Correspondingly, 
the phase matrix 
$P_{12} = \diag \,(1,1,1)$ in Eq. (\ref{eq:M2ddiag}).
Thus,  the unitary matrix  $U_d$ contributing to the CKM matrix, 
has the following simple form: 
%%%%%%%%%%%%%%%%%%%%%%%%%
\begin{equation}
 U_d = P_{23}\,O_{23}\,P_{13}\,O_{13}\,O_{12}\,,
\label{eq:Ud}
\end{equation}
%%%%%%%%%%%%%%%%%%
%
where the phases in the phase matrices 
$P_{23} = \diag\,(1,1,e^{-\,i\phi_{23}})$ and
$P_{13} = \diag\,(1,1,e^{-\,i(\phi_{13}-\phi_{23})})$
are determined by the phase of $g_d$:
%%%%%%%%%%%%%%%%%%%%%%%%%
\begin{equation}
\phi_{23} = {\rm arg}[g_d]\,,\qquad 
\phi_{13}= {\rm arg}\left [\dfrac{7}{\sqrt{37}}+\dfrac{3}{g^*_d}\right ]\,. 
\label{eq:ph23ph13}
\end{equation}
%%%%%%%%%%%%%%%%%%%%%%%%%%%
%

It follows from our results  that $\sin\theta^d_{12}$
depends on $|q_4|$ and  $|\hat\beta_d|/|\hat\alpha_d|$, 
while $\sin\theta^d_{23}$ and  
$\sin\theta^d_{13}$ depend only on $|q_4|$ and  
$g_d$. Thus, all  the three mixing angles are independent 
of the constant ratio
 $|\hat\alpha_d|/|g_d\hat\gamma_d|$.

%%%%%%%%%%%%%%%%%%%%%%%%%%%%%%%
%
\section{Mixing from the Up-type Quarks}
\label{sec:umixing}
%
%%%%%%%%%%%%%%%%%%%%%%%%%%%%%%%

Under the conditions specified in the preceding subsection 
(keeping terms not smaller than $|q_4|^4$, setting 
$\re[\tau] = 0$, $|q_4| \ll 1$,  
$|\hat\alpha_u|/|\hat\beta_u| \gg 1$ and 
$|\hat\gamma_u|/|\hat\beta_u \gg 1$), it is possible to diagonalise 
the real symmetric matrix $M^{(2)}_u$ and obtain analytic expression for the 
orthogonal matrix $O_u$ contributing to the CKM matrix:
%%%%%%%%%%%%%%%%%%%%%%%%%%%%%%%%
\begin{eqnarray}
\nonumber
&& M^{(2)}_u =  O_u\,M^{(2,diag)}_u\,O^T_u\,,\\[0.25cm]
&&
 O_u = O^u_{23}\,O^u_{13}\,O^u_{12}\,,
\label{eq:M2udiag}
\end{eqnarray}
%%%%%%%%%%%%%%%%%%
%
where $M^{(2,diag)}_d = \diag\,(m^2_u,m^2_c,m^2_t)$, 
and $O^u_{23}$, $O^u_{13}$ and $O^u_{12}$ are orthogonal matrices 
describing 2-3, 1-3 and 1-2 rotations. They have the same form as the 
matrices given in Eq.(\ref{eq:Oij}) with the angles 
replaced respectively by
$\theta^u_{23}$, $\theta^u_{13}$ and $\theta^u_{12}$.
%%%%%%%%%%%%%%%%%%%%%%%%%
% \begin{eqnarray}
% O_{23} = \begin{pmatrix}
% 1 & 0 & 0 \\
% 0 & c^d_{23} & s^d_{23} \\
% 0 &-\,s^d_{23}  & c^d_{23}
% \end{pmatrix}\,,\quad
% O_{13}= \begin{pmatrix}
% c^d_{13} & 0 & s^d_{13} \\
% 0 & 1 & 0 \\
% -\,s^d_{13} & 0  & c^d_{13}
% \end{pmatrix}\,,\quad
% O_{12}= \begin{pmatrix}
% c^d_{12} & s^d_{12} & 0 \\
% -\,s^d_{12} & c^d_{12} & 0 \\
% 0 & 0  & 1
% \end{pmatrix}\,. 
% \label{eq:Oij}
% \end{eqnarray}
%%%%%%%%%%%%%%%%%%%%%%%%%%
%
% Here $c^d_{ij} = \cos\theta^d_{ij}$ 
% and $s^d_{ij} = \sin\theta^d_{ij}$ 
% with $ij = 23,13,12$. 
Given the hierarchy of the elements of 
$M^{(2)}_u$, the angles $\theta^u_{ij}$ are determined by:
%%%%%%%%%%%%%%%%%%%%%%%%%%
\begin{eqnarray}
\nonumber
&&\tan2\theta^u_{23} = 
\dfrac{2(M^{(2)}_u)_{23}}{(M^{(2)}_u)_{33} - (M^{(2)}_u)_{22}}\,,\\[0.3cm]
\nonumber
&&\tan2\theta^u_{13} = 
\dfrac{2(M^{(2)}_u)_{13}}{(M^{(2)}_u)_{33} - (M^{(2)}_u)_{11}}\,,\\[0.3cm]
&&\tan2\theta^u_{12} = 
\dfrac{2(\tilde{M}^{(2)}_u)_{12}}{(\tilde{M}^{(2)}_u)_{22} - 
(\tilde{M}^{(2)}_u)_{11}}\,,
\label{eq:tan2thuij} 
\end{eqnarray}
%%%%%%%%%%%%%%%%%%%%%%%%%
%
where $(\tilde{M}^{(2)}_u)_{12}$,
$(\tilde{M}^{(2)}_u)_{22}$ and 
$(\tilde{M}^{(2)}_u)_{11}$ include corrections in the $(M^{(2)}_u)_{12}$, 
$(M^{(2)}_u)_{22}$ and 
$(M^{(2)}_u)_{11}$ elements due to the 2-3 and 1-3 rotations, respectively: 
%%%%%%%%%%%%%%%%%%%%%%%%%%%
\begin{eqnarray}
\nonumber
&& (\tilde{M}^{(2)}_u)_{12} = (M^{(2)}_u)_{12}\cos\theta^u_{23} - 
(M^{(2)}_u)_{13}\,\sin\theta^u_{23} 
\cong 
(M^{(2)}_u)_{12} - (M^{(2)}_u)_{13}\,\sin\theta^u_{23}\,,
\\[0.3cm]
\nonumber
&&(\tilde{M}^{(2)}_u)_{22} = (M^{(2)}_u)_{22}
\left (1 - \dfrac{|(M^{(2)}_u)_{23}|^2}
{ (M^{(2)}_u)_{22} ( (M^{(2)}_u)_{33} - (M^{(2)}_u)_{22} ) } \right) \,,\\[0.3cm]
&&(\tilde{M}^{(2)}_u)_{11} = (M^{(2)}_u)_{11}
\left (1 - \dfrac{((M^{(2)}_u)_{13})^2}
{(M^{(2)}_u)_{11}(M^{(2)}_u)_{33}} \right) \,.
\label{eq:tMu22tMu11}
\end{eqnarray}
%%%%%%%%%%%%%%%%%%%%%%%%%%
%
Given the values of $|\hat\alpha_u|/|\hat\beta_u|$ and 
$|\hat\gamma_u|/|\hat\beta_u|$ in Eq.(\ref{eq:u1}), which allow to explain 
the up-type quark mass hierarchies, 
the fact that $|q_4|\sim 10^{-2}$ and  
taking into account the expressions for 
the elements of  $M^{(2)}_u$ given in Eq. (\ref{eq:MudegMu})
it is not difficult to convince oneself 
that $(M^{(2)}_u)_{33} \gg (M^{(2)}_u)_{11},(M^{(2)}_u)_{22}$.
For $\sin\theta^u_{23}$ and  $\sin\theta^u_{13}$ we find:
%%%%%%%%%%%%%%%%%%%%%%%%%%%%%
\begin{eqnarray}
\label{eq:thu23}
&& \sin\theta^u_{23} \cong 12\,|q_4|^2\,,\\[0.25cm]
\label{eq:thu13}
&& \sin\theta^u_{13} \cong 64\,\sqrt{2}\,|q_4|^3\,.
\end{eqnarray}
%%%%%%%%%%%%%%%%%%%%%%%%
%
For the ``ingredients'' used for the calculation of 
$\sin\theta^u_{12}$ we get:
%%%%%%%%%%%%%%%%%%%%%%%%%%%%%%%%%
\begin{eqnarray}
\nonumber
&& \dfrac{(M^{(2)}_u)_{23}}{ (M^{(2)}_u)_{33} - (M^{(2)}_u)_{22}}
\cong 12\,|q_4|^2\,,
\nonumber\\[0.3cm]
&&\dfrac{(M^{(2)}_u)_{23}}{(M^{(2)}_u)_{22}}\cong 
\dfrac{1}{12\,|q_4|^2\,(1 + Z)}\,,\\[0.3cm]
\nonumber
&& (\tilde{M}^{(2)}_u)_{22} \cong (M^{(2)}_u)_{22}\,\dfrac{Z}{1+Z}\,,\\[0.3cm]
\nonumber
&& (\tilde{M}^{(2)}_u)_{11} = (M^{(2)}_u)_{11}\,
\dfrac{9}{64}\,Z \dfrac{1}{1 + \dfrac{9}{64}Z}\,,\\[0.3cm]
\nonumber
&& \dfrac{ (M^{(2)}_u)_{11}}{M^{(2)}_u)_{22}} \cong 
8\,|q_4|^2\,\dfrac{64}{9}\,\dfrac{1 + \dfrac{9}{64}Z}{1+Z}\,,\\[0.3cm]
\nonumber
&& (\tilde{M}^{(2)}_u)_{22} - (\tilde{M}^{(2)}_u)_{11} = 
\dfrac{Z}{1 + Z}\,M^{(2)}_u)_{22}( 1 - 8|q_4|^2)\,,\\
%\label{eq:tMu22mintMu11}
%\end{eqnarray}
%%%%%%%%%%%%%%%%%%%%%%%%
%\begin{eqnarray}
\nonumber
&& \dfrac{ (M^{(2)}_u)_{13}}{ (M^{(2)}_u)_{12}} \cong 
\dfrac{1}{ 12\,|q_4|^2\left (1 - \dfrac{3}{8}\,Z\right)}\,,\\[0.3cm]
\nonumber
&& (\tilde{M}^{(2)}_u)_{12} \cong 
(M^{(2)}_u)_{12}
\left (1 - \dfrac{\sin\theta^u_{23}}
{12\,|q_4|^2 (1 - \dfrac{3}{8}\,Z) }
\right)
\cong (M^{(2)}_u)_{12} \dfrac{-\,\dfrac{3}{8}\,Z}
{1 - \dfrac{3}{8}\,Z}\,, 
\nonumber
\\[0.3cm]
&& \dfrac{(M^{(2)}_u)_{12}}{(M^{(2)}_u)_{22}} \cong
\dfrac{16\sqrt{2}}{3}\,|q_4|\,\dfrac{1 - \dfrac{3}{8}\,Z}{1 + Z}\,,
\label{eq:ingrthu12}
\end{eqnarray}
%%%%%%%%%%%%%%%%%%%%%%%%%%%%%%%%
%
where 
%%%%%%%%%%%%%%%%%%%%%%%%
\begin{equation}
Z = |q_4|^{-2}\,\dfrac{|\hat\alpha_u|}{|\hat\gamma_u|}\,,
\label{eq:Z}
\end{equation}
%%%%%%%%%%%%%%%%%%%%%%%%%%%
%
and we have used the fact that $\sin\theta^u_{23} = 12\,|q_4|^2$ 
\footnote{We note that 
for $|q_4|=2.7402\times 10^{-2}$,
$|\hat\alpha_u|/|\hat\beta_u| = 18.862$ 
and $|\hat\gamma_u|/|\hat\beta_u| = 2194$,
which allow to reproduce the up-type quark mass hierarchies, 
we have $Z = 0.0966$.
}.
Combining the results on the ``ingredients'', which 
are part of the calculation of $\sin\theta^u_{12}$, we get the 
following simple result:
%%%%%%%%%%%%%%%%%%%%%%%%%%%5
\begin{equation}
\sin\theta^u_{12} \cong -\,2\sqrt{2}\,|q_4| + O(10^{-4})\,.
\label{eq:thu12}
\end{equation}
%%%%%%%%%%%%%%%%%%%%%%%%%%
%

For the Cabibbo angle, as we will show below, we have:
%%%%%%%%%%%%%%%%%%%%%%%%%%%5
\begin{equation}
\sin\theta_{12} = \sin\theta^d_{12} - (\sin\theta^u_{12}) 
= \sin\theta^d_{12} + |\sin\theta^u_{12}|\,.
\label{eq:th12}
\end{equation}
%%%%%%%%%%%%%%%%%%%%
%

%%%%%%%%%%%%%%%%%%%%%%%%%%%%%%%
%
\section{The CKM Matrix}
\label{sec:CKM}
%
%%%%%%%%%%%%%%%%%%%%%%%%%%%%%%%
%

In the standard parametrisation the CKM matrix has the form:
%%%%%%%%%%%%%%%%%%%%%%%%%%%%%%%%
\begin{eqnarray}
&&V_{\rm CKM} =
\left(\begin{array}{ccc}
V_{ud} & V_{us}  & V_{ub}  \\
V_{cd}  & V_{cs}  & V_{cb}  \\
V_{td} & V_{ts}  & V_{tb}
\end{array}\right)
\nonumber\\
%\\[0.25cm}
&&\hskip 1.15 cm = \left(\begin{array}{ccc}
c_{12}c_{13} & s_{12}c_{13} & s_{13}e^{-i\delta}\\
-s_{12}c_{23}-c_{12}s_{13}s_{23}e^{i\delta_{CP}} & c_{12}c_{23}-s_{12}s_{13}s_{23}e^{i\delta}  & c_{13}s_{23}\\
s_{12}s_{23}-c_{12}s_{13}c_{23}e^{i\delta} & -c_{12}s_{23}-s_{12}s_{13}c_{23}e^{i\delta} 
&  c_{13}c_{23}
\end{array}\right)\,,
\label{eq:VCKM}
\end{eqnarray}
%%%%%%%%%%%%%%%%%%%%%%%%%%%%%%%%%
%
where $c_{ij}\equiv \cos\theta_{ij}$, $s_{ij}\equiv \sin\theta_{ij}$ and $\delta$ 
is the CP violation (CPV) phase.
 The angles $\theta_{ij}$ and the phase 
$\delta$ have been experimentally determined 
with a relatively high precision.
Thus, as is well known, to a good approximation, 
$V_{ud} \cong c_{12}$, $V_{us} \cong s_{12}$,
$V_{cb} \cong  s_{23}$ and $V_{tb} \cong  c_{23}$.
Since the quark flavour model considered by us is given at 
the GUT scale of  $M_{\rm GUT} = 2\times 10^{16}$ GeV, we use 
the numerical values of $\theta_{ij}$ and $\delta$ at the GUT scale 
obtained by the RG evolution.

  In the model constructed by us, the CKM matrix is given by:
%%%%%%%%%%%%%%%%%%%%%%%%%%%%%%
\begin{eqnarray}
\nonumber
&& U_{\rm CKM} = O^T_u\,U_d
= \left(\begin{array}{ccc}
U_{ud} & U_{us}  & U_{ub}  \\
U_{cd}  & U_{cs}  & U_{cb}  \\
U_{td} & U_{ts}  & U_{tb}
\end{array}\right)
\,,\\[0.25cm]
\nonumber
&& O_u = O^u_{23}\,O^u_{13}\,O^u_{12}\,, \qquad
U_d = P_{23}\,O^d_{23}\,P_{13}\,O^d_{13}\,O^d_{12}\,,\\[0.25cm]
&& P_{23} = \diag\,(1,1,e^{-\,i\phi_{23}})\,,\qquad
P_{13} = \diag\,(1,1,e^{-\,i(\phi_{13}-\phi_{23})})\,,
\label{eq:UCKM1}
\end{eqnarray}
%%%%%%%%%%%%%%%%%%%%%%%
%
where the phases 
% $\phi_{23} = arg(g_d)$ and 
% $\phi_{13} = arg\left (\dfrac{7}{\sqrt{37}} + \dfrac{3}{g^*_d} \right )$ 
$\phi_{23}$ and $\phi_{13}$ are defined in Eq. (\ref{eq:ph23ph13}) and
we have used the notation  $U_{\rm CKM}$ instead of  $V_{\rm CKM}$
to indicate that this is the theoretically derived CKM matrix.
The matrix $O_u$ is given by:
%%%%%%%%%%%%%%%%%%%%%%%%%%%%%
\begin{equation}
 O_u = 
\left(\begin{array}{ccc}
c^u_{12}c^u_{13}  &   s^u_{12}c^u_{13}   &   s^u_{13}  \\
-\,s^u_{12}c^u_{23} -\,c^u_{12}s^u_{13}s^u_{23}e & c^u_{12}c^u_{23} 
-\,s^u_{12}s^u_{13}s^u_{23}  &  c^u_{13}s^u_{23}  \\
s^u_{12}s^u_{23} -\,c^u_{12}s^u_{13}c^u_{23}  & -\,c^u_{12}s^u_{23} 
-\,s^u_{12}s^u_{13}c^u_{23} &  c^u_{13}c^u_{23}
\end{array}\right)\,.
\label{eq:Ou}
\end{equation}
%%%%%%%%%%%%%%%%%%%%%%%%%%
%

It proves convenient to cast the matrix $U_d$ in the form:
%%%%%%%%%%%%%%%%%%%%%%%%%
\begin{equation}  
U_d = P_{23}\,O^d_{23}\,P_{13}\,O^d_{13}\,O^d_{12}
= U^d_{23}\,U^d_{13}\,O^d_{12}\,P^\prime_{13}\,,
\label{eq:Ud2}
\end{equation}
%%%%%%%%%%%%%%%%%%%%%%%%%%%%%%
%
where $P^\prime_{13} = \diag\,(1,1,e^{-\,i\phi_{13}})$ and
%%%%%%%%%%%%%%%%%%%%%%%%
\begin{eqnarray}
U^d_{23} = \begin{pmatrix}
1 & 0 & 0 \\
0 & c^d_{23} & s^d_{23}\,e^{i\phi_{23}} \\
0 &-\,s^d_{23}\,e^{-i\phi_{23}}  & c^d_{23}
\end{pmatrix}\,,\qquad
U^d_{13}= \begin{pmatrix}
c^d_{13} & 0 & s^d_{13}\,e^{i\phi_{13}} \\
0 & 1 & 0 \\
-\,s^d_{13}\,e^{-i\phi_{13}} & 0  & c^d_{13}
\end{pmatrix}\,.
% P^\prime_{13} = diag(1,1,e^{-\,i\phi_{13}})\,.
\label{eq:Ud23Ud13}
\end{eqnarray}
%%%%%%%%%%%%%%%%%%%%%%%%%%%
%
Thus, for the matrix $U^d$ we have: 
%%%%%%%%%%%%%%%%%%%%%%%%%%%%%%%%%%%%%%%%
\begin{equation}
\label{eq:Ud3}
U_d= P_{23}\,\left(\begin{array}{ccc}
c^d_{12}c^d_{13} & s^d_{12}c^d_{13} & s^d_{13}e^{-i\phi}\\
-s^d_{12}c^d_{23}-c^d_{12}s^d_{13}s^d_{23}e^{i\phi} 
& c^d_{12}c^d_{23}-s^d_{12}s^d_{13}s^d_{23}e^{i\phi} & c^d_{13}s^d_{23}  \\
s^d_{12}s^d_{23}-c^d_{12}s^d_{13}c^d_{23}e^{i\phi} 
&-c^d_{12}s^d_{23}-s^d_{12}s^d_{13}c^d_{23}e^{i\phi} & c^d_{13}c^d_{23}
\end{array}\right)\,P^*_{23}P^\prime_{13}\,,
% ~\phi = \phi_{23}-\phi_{13}\,.
\end{equation}
%%%%%%%%%%%%%%%%%%%%%%%%%%%%%%%%%
%
where $\phi = \phi_{23}-\phi_{13}$.
The phase matrix $P^*_{23}P^\prime_{13}$ 
can be absorbed by the left-handed down-type  quark fields.
Using Eqs.(\ref{eq:UCKM1}), (\ref{eq:Ou}) and 
(\ref{eq:Ud3}) we get the following analytic 
expressions for the elements of the first row and third column 
of $U_{CKM}$:
%%%%%%%%%%%%%%%%%%%%%%%%%%%%%%%%%%
\begin{eqnarray}
\nonumber
&& U_{ud} = \cos(\theta^d_{12} - \theta^u_{12}) + O(10^{-5}) = 
\cos\theta_{12}\,,\\[0.25cm]
\nonumber
&& U_{us} = \sin(\theta^d_{12} - \theta^u_{12}) + O(10^{-4}) = 
\sin\theta_{12}\,,\\[0.25cm]
\nonumber
&& U_{ub} = e^{-i(\phi_{23} + \kappa_{cb})}
\left ((s^d_{13}\,e^{i\phi_{13}} - s^u_{13})\,e^{i\kappa_{cb}} -
s^u_{12}\left |s^d_{23}\,e^{i\phi_{23}} - s^u_{23} \right|
\right ) + O(10^{-5})\,,\\[0.25cm]
\nonumber
&& U_{cb} = e^{-i\phi_{23}}\,\left (e^{i\phi_{23}}s^d_{23}c^u_{23} 
- c^d_{23}s^u_{23}\right ) + O(10^{-4})
\\[0.25cm]
\nonumber
&& \hskip 0.6 cm \cong e^{-i(\phi_{23} + \kappa_{cb})}
\left|s^d_{23}\,e^{i\phi_{23}} - s^u_{23}\right| +  O(10^{-4}) 
\cong  e^{-i(\phi_{23} + \kappa_{cb})}\,\sin\theta_{23}
\,,\\[0.25cm] 
&& U_{tb} = e^{-i\phi_{23}}\,\left (e^{i\phi_{23}}s^d_{23}s^u_{23} 
+ c^d_{23}c^u_{23}\right ) + O(10^{-5})
\cong e^{-i\phi_{23}}\,\cos\theta_{23}\,,
\label{eq:UuiUcbUct}
\end{eqnarray}
%%%%%%%%%%%%%%%%%%%%%%%%%%%%%
%
where 
%%%%%%%%%%%%%%%%%%%%%%
\begin{eqnarray}
\nonumber
&& |U_{cb}| = 
\left ( \sin^2(\theta^d_{23} - \theta^u_{23}) 
+ 2\,s^d_{23}\,s^u_{23}\left ( 1 - \cos\phi_{23}\right)
\right )^{\frac{1}{2}} \cong \sin\theta_{23}\,,\\[0.25cm]
\nonumber
&& \sin\kappa_{cb} = -\,\dfrac{s^d_{23}\,\sin\phi_{23}}{|U_{cb}|}\,,
\quad \cos\kappa_{cb} = \dfrac{s^d_{23}\,\cos\phi_{23}-s^u_{23}}{|U_{cb}|}\,,
\\[0.25cm]
&& |U_{tb}| \cong \left ( \cos^2(\theta^d_{23} - \theta^u_{23}) 
- 2\,s^d_{23}\,s^u_{23}\left ( 1 - \cos\phi_{23}\right)
\right )^{\frac{1}{2}} \cong \cos\theta_{23}\,. 
\label{eq:UcbUtb}
\end{eqnarray}
%%%%%%%%%%%%%%%%%%%%%%%%%%%
%
In Eq. (\ref{eq:UuiUcbUct}),  
$\theta_{12}$ and $\theta_{23}$ are the CKM angles.
As we will show below, the phase factors
$ e^{-i(\phi_{23} + \kappa_{cb})}$ and  $e^{-i\phi_{23}}$ 
in the elements $U_{ub}$, $U_{cb}$ and $U_{tb}$
do not play a role in the CKM mixing phenomenology.
Thus, it is convenient to cast  $U_{ub}$ in the form
$U_{ub} = e^{-i(\phi_{23} + \kappa_{cb})}\tilde{U}_{ub}$ with
%%%%%%%%%%%%%%%%%%%%%%%%%%%%%%%%%
\begin{equation}
 \tilde{U}_{ub} = \sin\theta_{13}\,e^{-i\,\delta}\,,
\label{eq:tUtb1}
\end{equation}
%%%%%%%%%%%%%%%%%%%%%%%%%%%%%
% 
$\theta_{13}$ and $\delta$ being the CKM 
mixing angle and CPV phase,
%%%%%%%%%%%%%%%%%%%%%%%%%%%
\begin{eqnarray}
\nonumber
&& \sin\theta_{13} = |\tilde{U}_{ub}| 
\cong \left| (s^d_{13}\,e^{i\phi_{13}} - s^u_{13})\,e^{i\kappa_{cb}} -
s^u_{12}\left |s^d_{23}\,e^{i\phi_{23}} - s^u_{23} \right|
\right|\,,\\[0.25cm]
&& \delta =
{\rm arg}\left[(s^d_{13}\,e^{-i\phi_{13}} - s^u_{13})\,e^{-i\kappa_{cb}} -
s^u_{12}\left |s^d_{23}\,e^{-i\phi_{23}} - s^u_{23} \right|
\right]\,.
\label{eq:s13d}
\end{eqnarray}
%%%%%%%%%%%%%%%%%%%%%%%
%

The elements $U_{cs}$, $U_{td}$ and $U_{ts}$ and given by:
%%%%%%%%%%%%%%%%%%%%%%%%%%%
\begin{eqnarray}
\nonumber
&& U_{cs} = \cos(\theta^d_{12}-\theta^u_{12})\, \cos(\theta^d_{23}-\theta^u_{23}) + 
O(10^{-4})\,,\\[0.25cm]
\nonumber
&& U_{td} \cong e^{i\kappa_{cb}} 
\left (
(s^d_{12}-s^u_{12})\,
\left |s^d_{23}\,e^{i\phi_{23}} - s^u_{23} \right| - 
\left(
(s^d_{13}\,e^{-i\phi_{13}} - s^u_{13})\,e^{-i\kappa_{cb}} -
s^u_{12}\left (s^d_{23}\,e^{-i\phi_{23}} - s^u_{23} \right|
\right )
\right)\\[0.25cm]
\nonumber
&&\hskip 0.6 cm \cong  e^{i\kappa_{cb}}
\left (
\sin\theta_{12}\,\sin\theta_{23} -\, 
\sin\theta_{13}\,e^{i\delta}
\right )\,,\\[0.25cm]
&&  U_{ts} \cong -\,e^{i\kappa_{cb}}\left|s^d_{23}\,e^{-i\phi_{23}} - s^u_{23} \right| 
+ O(s_{12}s_{13}) \cong -\,e^{i\kappa_{cb}}\,\sin\theta_{23}\,.
\label{eq:UcsUtdUts} 
\end{eqnarray}
%%%%%%%%%%%%%%%%%%%%%%%%%
%
We extract the common phase of $e^{i\kappa_{cb}}$ of 
$U_{td}$ and $U_{ts}$ on the left in a diagonal phase matrix 
$P_{cb} = \diag(1,1,e^{i\kappa_{cb}})$, which can be absorbed 
by the $t$-quark field. As a consequence,
the element $U_{tb}$ acquires a phase  $e^{-i\kappa_{cb}}$.
However, together with the phase $e^{-i\phi_{23}}$, 
$e^{-i\kappa_{cb}}$ becomes a common phase of the last column of 
$U_{\rm CKM}$ and can be factorised on the right in a phase matrix 
$P_{cb+23} = \diag(1,1,e^{-i(\phi_{23}+\kappa_{cb})})$, 
which can in turn (together with the phase matrices 
$P^*_{23}P^\prime_{13}$)
be absorbed by the $b$-quark field.
As a result, $U_{\rm CKM}$ takes the forms of the standard parametrisation of the 
CKM matrix given in Eq. (\ref{eq:VCKM}).

We summarise below the analytic 
expressions for the CKM angles and the CPV phase $\delta$ 
in the model we are considering:
%%%%%%%%%%%%%%%%%%%%%%%%%%%%%%%%5
\begin{eqnarray}
\nonumber
&& \cos\theta_{12} \cong \cos(\theta^d_{12} - \theta^u_{12})\,,
\quad \sin\theta_{12} \cong \sin(\theta^d_{12} - \theta^u_{12})\,,
\quad  \theta^u_{12} < 0\,,\\[0.25cm]
\nonumber
&& \sin\theta_{23} \cong 
\left ( \sin^2(\theta^d_{23} - \theta^u_{23}) 
+ 2\,s^d_{23}\,s^u_{23}\left ( 1 - \cos\phi_{23}\right)
\right )^{\frac{1}{2}}\,,\\[0.25cm]
\nonumber
&&  \cos\theta_{23} \cong
\left ( \cos^2(\theta^d_{23} - \theta^u_{23}) 
- 2\,s^d_{23}\,s^u_{23}\left ( 1 - \cos\phi_{23}\right)
\right )^{\frac{1}{2}}\,,\\[0.25cm]
\nonumber
&& 
\sin\theta_{13}
\cong \left| (s^d_{13}\,e^{i\phi_{13}} - s^u_{13})\,e^{i\kappa_{cb}} -
s^u_{12}\left |s^d_{23}\,e^{i\phi_{23}} - s^u_{23} \right|
\right|\,,\\[0.25cm]
&& \delta =
\arg\left[(s^d_{13}\,e^{-i\phi_{13}} - s^u_{13})\,e^{-i\kappa_{cb}} -
s^u_{12}\left |s^d_{23}\,e^{-i\phi_{23}} - s^u_{23} \right|
\right]\,.
\label{eq:thijd}
\end{eqnarray}
%%%%%%%%%%%%%%%%%%%%%%%%%%%%%
%

From Eq. (\ref{eq:s13d}) we find:
%%%%%%%%%%%%%%%%%%%%%%%%%%%%%
\begin{equation}
\sin\theta_{13}\,\sin\delta \cong
 -\,s^d_{13}\,\sin(\phi_{13} + \kappa_{cb}) + s^u_{13}\sin\kappa_{cb}
= -\,\sin\phi_{23}\,|q_4|^5\,\dfrac{96\,\times 18\sqrt{74}}{|g_d|\,|U_{cb}|}\,,
\label{eq:s13sd}
\end{equation}
%%%%%%%%%%%%%%%%%%%%%%%%%%%%%%%%
%
where the second equation was obtained by employing
 the expressions  for $\sin\phi_{13}$,  $\cos\phi_{13}$, 
$\sin\kappa_{cb}$ and $\cos\kappa_{cb}$ 
given in Eqs. (\ref{eq:ph23ph13}) and (\ref{eq:UcbUtb}).
By sybstituting the analytic results for 
$s^d_{13}$, $s^u_{13}$, $\sin\kappa_{cb}$ and $\cos\kappa_{cb}$,
$s^u_{12}$, $s^d_{23}$ and $s^u_{23}$, reported 
in Eqs. (\ref{eq:sth13d}),  (\ref{eq:thu13}),  (\ref{eq:UcbUtb}),
(\ref{eq:thu12}), (\ref{eq:sth23d}) and (\ref{eq:thu23})
in the expression for $\sin\theta_{13}$ in Eq. (\ref{eq:thijd})
we find a rather simple expression for $\sin\theta_{13}$:
% {eq:UcbUtb}{eq:thu12}{eq:thu13}{eq:thu23}{eq:sth13d}l{eq:sth23d}
%%%%%%%%%%%%%%%%%%%%%%%%%%
\begin{equation}
\sin\theta_{13} = 144\,\sqrt{2}\, |q_4|^3\,.% + O(10^{-5}). 
\label{eq:th13}
\end{equation}
%%%%%%%%%%%%%%%%%%%%%%%%%%%
%
Correspondingly, for $\sin\delta$ we get from Eq. (\ref{eq:s13sd}):
%%%%%%%%%%%%%%%%%%%%%%%%%%
\begin{equation}
\sin\delta = -\,\sin\phi_{23}\,|q_4|^2\,\dfrac{12\,\sqrt{37}}
{|g_d|\,\sin\theta_{23}}\,.
\label{eq:sd}
\end{equation}
%%%%%%%%%%%%%%%%%%%%%%%%%
%

We recall further that $|U_{cb}| = |s^d_{23}\,e^{i\phi_{23}} - s^u_{23}| 
\cong \sin\theta_{23}$. Using the results on 
$s^d_{23}$ and $s^u_{23}$ from Eqs. (\ref{eq:sth23d}) and 
(\ref{eq:thu23}) and given the value of $\sin\theta_{23}$,  
it is possible to derive the correlation between 
$|g_d|$ and $\cos\phi_{23}$ which allows to reproduce the value 
of $\sin\theta_{23}$:
%%%%%%%%%%%%%%%%%%%%%%%%%%%%%%
\begin{equation}
\dfrac{\sqrt{37}}{|g_d|} = 
\cos\phi_{23} + 
\left (\cos^2\phi_{23} + 
\frac{\sin^2\theta_{23}}{(12|q_4|^2)^2} - 1\right)^{\frac{1}{2}}\,.
\label{eq:gdc23}
\end{equation}
%%%%%%%%%%%%%%%%%%%%%%%%%%%
%
The minimal (maximal) value of  $|g_d|$ corresponds to 
$\cos\phi_{23} = 1$ ($\cos\phi_{23} = 0$), 
i.e., to CP-symmetry conserving (CP-symmetry violating) $g_d$.
For, e.g., $|q_4| = 2.7402\times 10^{-2}$ and $\sin\theta_{23} = 0.04$ 
we get $|g_d| \cong 1.118$ ($|g_d| \cong 1.406$).

 Finally,  for the $J_{CP}$ factor \cite{Jarlskog:1985ht} we find:
%%%%%%%%%%%%%%%%%%%%%%%%%%%%%%%%
\begin{eqnarray}
\nonumber
&& J_{CP} = \im\left[ U_{us}\,U_{cb}\,U^*_{ub}\,U^*_{cs}\right]\\[0.25cm] 
\nonumber
&& \hskip 0.7 cm \cong 
\sin(\theta^d_{12} - \theta^u_{12})\,
|s^d_{23}\,e^{i\phi_{23}} - s^u_{23}|\,
\left(-\,s^d_{13}\,\sin(\phi_{13} + \kappa_{cb}) 
+ s^u_{13}\,\sin\kappa_{cb}\right )
 \\[0.25cm]
&& \hskip 0.7 cm = \sin\theta_{12}\, \sin\theta_{23}\,\sin\theta_{13}\sin\delta\,,
\label{eq:JCP1}
\end{eqnarray}
%%%%%%%%%%%%%%%%%%%%%%%%%
%
where we have used Eqs. (\ref{eq:thijd}) and (\ref{eq:s13sd}).
Substituting $\sin\theta_{13}\sin\delta$ with the expression in 
Eq. (\ref{eq:s13sd}) and taking into account that 
$\sin\theta_{23} \cong |U_{cb}| \neq 0$ we obtain:
%%%%%%%%%%%%%%%%%%%%%%%%%
\begin{equation} 
J_{CP} \cong -\,\sin(\theta^d_{12} - \theta^u_{12})\,
\sin\phi_{23}\,|q_4|^5\,\dfrac{96\times 18\sqrt{74}}{|g_d|}\,.
\label{eq:Jcp}
\end{equation}
%%%%%%%%%%%%%%%%%%%%%%%%%%%%
% 

%%%%%%%%%%%%%%%%%%%%%%%%%%%%%%%
%
\section{Numerical Analysis}
\label{sec:numerics}
%
%%%%%%%%%%%%%%%%%%%%%%%%%%%%%%%

While the value of $\sin\theta_{12}$ (or $\theta_{12}$) 
has been measured with very high precision in low-energy experiments,
there exist ambiguities as to what the precise values of $|V_{cb}|$  
and $|V_{ub}|$ are. As is well known, to a very good approximation we have
in the standard parametrisation of the CKM matrix we are employing 
in the present study 
$|V_{cb}| \cong \sin\theta_{23} \cong \theta_{23}$  
and $|V_{ub}| \cong \sin\theta_{13} \cong \theta_{13}$.
There are two different sets of values of $\theta_{23}$ and $\theta_{13}$, 
obtained from the low-energy inclusive and exclusive decay data 
\footnote{For details see the review articles
``Semileptonic $b$-Hadon Decays, Determination of 
$V_{cb}$ and $V_{ub}$ by A.X. El-Khadra and P. Uruqijo, 
and  ``CKM Quark-Mixing Matrix''
 by A. Ceccucci, Z. Ligeti and Y. Sakai 
in \cite{PDG2024ElKU,PDG2024CLS}.}
\cite{PDG2024ElKU,PDG2024CLS}: from the inclusive decay data it is found that  
$\theta_{23} = (4.22 \pm 0.05)\times 10^{-2}$ and
$\theta_{13} = 
(4.13 \pm 0.12_{exp}\,(^{+0.13}_{-0.14})_{theo} \pm 0.18_{\delta mod})\times 10^{-3}$,
$(^{+0.13}_{-0.14})_{theo}$ and $\pm 0.18_{\delta mod}$ being the 
theoretical and model uncertainties,
while from the exclusive decay data it was obtained   
$\theta_{23} = (3.98 \pm 0.06)\times 10^{-2}$ and
$\theta_{13} = (3.70 \pm 0.10 \pm 0.12)\times 10^{-3}$.  
These are values at the electroweak scale $\sim M_Z$. Thus,   
the ``inclusive'' and ``exclusive''  values differ significantly. 
At present it is not clear what is the origin of the difference 
between these two sets of measured values of $\theta_{23}$ and $\theta_{13}$.
The problem is subject to debate and further studies.
If we add in quadratures the 
% statistical and systemtic 
errors in the results for  $\theta_{13}$ 
we obtain:
%%%%%%%%%%%%%%%%%%%%%%%%%%%%%%%%%
\begin{eqnarray}
\nonumber
&& \theta_{13} =  (4.13 ^{+0.252}_{-0.258})\times 10^{-3}\,, 
\quad\ \ \, \theta_{23} = (4.22 \pm 0.05)\times 10^{-2}\,,~~~
{\rm inclusive~data}\,,\\[0.25cm]
&& \theta_{13} =  (3.70 \pm 0.156)\times 10^{-3}\,,
\quad \theta_{23} = (3.98 \pm 0.06)\times 10^{-2}\,,
~~~{\rm exclusive~data}\,,
% \theta_{13} =  (3.44\pm 0.07)\times 10^-3\,,~~{\rm exclusive~data}\,,
\label{eq:th13th23atMz}
\end{eqnarray}
%%%%%%%%%%%%%%%%%%%%%%%%%%%%%%%%
%
where the error in the ``inclusive'' data value of $\theta_{13}$ corresponds 
% to $1\sigma$ error of
$+\sqrt{(0.12)^2 + (0.13)^2 + (0.18)^2}$
and $-\sqrt{(0.12)^2 + (0.14)^2 + (0.18)^2}$ 
\footnote{Any other treatment of the errors in the inclusive decay 
data on $\theta_{13}$ leads to a larger uncertainty in the value of 
$\theta_{13}$.
}, 
and we have added also the  ``inclusive'' and  ``exclusive''
values of $\theta_{23}$ for completeness.
In \cite{PDG2024CLS} the authors give also ``recommended  
average'' values of  $\theta_{23}$ and $\theta_{13}$:
%%%%%%%%%%%%%%%%%%%%%%%
\begin{equation}
\theta_{23} = (4.11 \pm 0.12)\times 10^{-2}\,, \quad
\theta_{13} = (3.82 \pm 0.20)\times 10^{-3}\,,
~~~~{\rm 2024~PDG~average~values}\,.
\label{eq:avth13th23atMz}
\end{equation}
%%%%%%%%%%%%%%%%%%%%%%%%%%%%%%%%%%
%

 In the following analysis we use the following input data 
at $M_{\rm GUT} = 2\times 10^{16}$ GeV, obtained by accounting for the RG 
% running 
effects on $\theta_{12}$, $\theta_{23}$ and 
$\theta_{13}$ for $\tan\beta = 10$ and SUSY breaking scale 
$M_{\rm SUSY} = 10$ TeV \cite{Antusch:2025fpm}:
%%%%%%%%%%%%%%%%%%%%%%%%%%%%%%%%%
\begin{eqnarray}
\label{eq:th12MGUT}
&&\hskip - 1 cm \theta_{12} = 0.226235 \pm 0.00085\,~\ 
(\sin\theta_{12} = 0.224310 \pm 0.00085)\,,
\\[0.25cm]
\label{eq:th23th13Msusy10TMGUTincl}
&&\hskip - 1 cm \theta_{23} = (3.939 \pm 0.0466 )\times 10^{-2}\,, 
\ \  \theta_{13} = (3.84^{+ 0.2347}_{-0.2396})\times 10^{-3}\,,
~~~~{\rm inclusive~data}\,,
\\[0.25cm]
\label{eq:th23th13MGUTexcl}
&&\hskip - 1 cm \theta_{23} =  (3.71 \pm 0.056)\times 10^-2\,,~
\ \ \ \, \theta_{13} =  (3.44 \pm 0.145)\times 10^{-3}\,,~~~{\rm exclusive~data}\,,
\\[0.25cm]
\label{eq:th23th13MGUTaver}
&& \hskip - 1 cm\theta_{23} =  (3.83 \pm 0.112)\times 10^{-2}\,, \ \
 \theta_{13} =  (3.55\pm 0.186)\times 10^{-3}\,,
~~{\rm 2024~PDG~average~values}\,.
\end{eqnarray}
%%%%%%%%%%%%%%%%%%%%%%%%%%%%%%%%
% 
For the Dirac CPV phase  $\delta$ we use the following value at the 
GUT scale:
%%%%%%%%%%%%%%%%%%%%%%%%%%%%
\begin{equation}
 \delta = (65.7^\circ \pm 3.0^\circ)\,.
\label{eq:dCPV}
\end{equation}
%%%%%%%%%%%%%%%%%%%%%%%%%%
%
The quark mass ratios at the GUT scale are presented in 
Eqs.(\ref{eq:mumt}), (\ref{eq:mcmt}), (\ref{eq:mdmb}) and 
(\ref{eq:msmb}). 

We will present results also in the case of high  
SUSY breaking scale $M_{\rm SUSY}$, considering 
the ``extreme'' case of 
$M_{\rm SUSY} = M_{\rm GUT}$. In this case
%%%%%%%%%%%%%%%%%%%%%%%%%%%%%%%
\footnote{
 In the case of high scale $M_{\rm SUSY}$ there is a RG 
contribution to the SM Higgs particle mass $m_\text{h}$, which depends on 
$M_{\rm SUSY}$, $\tan\beta$ and the mass of the top 
quark $m_t$ \cite{Giudice:2011cg}. 
Using the measured values of $m_\text{h}$ and $m_\text{t}$ 
one can obtain an upper limit on the possible values of 
$M_{\rm SUSY}$ for which the RG effects on  $m_\text{h}$   
are compatible with the experimentally determined  $m_\text{h}$. 
For relatively small values of $\tan\beta\sim 1$ and 
accounting for the $3\sigma$ uncertainties in the 
measured values of  $m_\text{h}$ and  $m_\text{t}$, the resulting upper 
limit on  $M_{\rm SUSY}$ is very loose and values of 
$M_{\rm SUSY}\sim 10^{16}$ GeV cannot be ruled out 
\cite{Bagnaschi:2014rsa}.
}
%%%%%%%%%%%%%%%%%%%%%%%%%%%%%%%

the proper SUSY effects in the RG evolution of the 
quark masses and mixing angles are negligible 
and these observables change
with the scale as in the Standard Model.  
Thus, at $M_{\rm GUT}$ we have \cite{Antusch:2025fpm}:
%%%%%%%%%%%%%%%%%%%%%%%%%%%%%%%%%
\begin{eqnarray}
\nonumber
&& \dfrac{m_d}{m_b} = (1.07\pm0.27)\times 10^{-3} \,, \quad\ \  \ 
\dfrac{m_s}{m_b} = (2.130 \pm 0.054)\times 10^{-2}\,,
\\[0.25cm]
\nonumber
&& \dfrac{m_u}{m_t} = (6.440 \pm 0.227)\times 10^{-6}\,, \quad 
\dfrac{m_c}{m_t} = (3.26 \pm 0.10)\times 10^{-3}\,,
\\[0.25cm]
\nonumber
%\label{eq:th12MGUT}
&& \theta_{12} = 0.226235 \pm 0.00085\,~
(\sin\theta_{12} = 0.224310 \pm 0.00085)\,,
\\[0.25cm]
% \label{eq:th23th13Msusy10TMGUTincl}
&& \theta_{23} = (4.65 \pm 0.136 )\times 10^{-2}\,, 
~ \theta_{13} = (4.326 \pm 0.226  )\times 10^{-3}\,,{\rm~PDG~average~values }\,,
\label{eq:atMGUTSM}
\end{eqnarray}
%%%%%%%%%%%%%%%%%%%%%%%%%%%%%%%%
%
where we have quoted only the ``average'' values of 
$\theta_{23}$ and alues $\theta_{13}$.

We present below the results of the statistical analyses of 
the three cases of experimental values of $\theta_{23}$ and 
$\theta_{13}$, quoted in 
Eqs. (\ref{eq:th23th13Msusy10TMGUTincl})-(\ref{eq:th23th13MGUTaver})
with $\tan\beta = 10$ and $M_{\rm SUSY} = 10$ TeV, and in the case of 
the ``average''   $\theta_{23}$ and 
$\theta_{13}$ for $M_{\rm SUSY} = M_{\rm GUT}$, 
given in Eq. (\ref{eq:atMGUTSM}). 
The experimental values of the quark mass 
ratios and $\theta_{12}$ in the last case are also given in  
Eq. (\ref{eq:atMGUTSM}), while the experimental value 
of $\delta$ is the same as that shown in Eq. (\ref{eq:dCPV}).

As a measure of goodness of fit in our statistical analysis, we use the sum of one-dimensional 
$\Delta\chi^2$ for eight   observable quantities
$O_j=(m_d/m_b, \, m_s/m_b,\,m_u/m_t, \, m_c/m_t,\, \theta_{12},\, \theta_{23},\, \theta_{13},\, \delta)$.
By employing the Gaussian approximation, we define
$\Delta \chi^2$:
%%%%%%%%%%%%%%%%%%%%%%%
\begin{align}
\Delta \chi^2=\sum_i\Delta \chi^2_i\,,\qquad 
\Delta \chi^2_i=\left ( \frac{O_j-O_{j,{\rm best\,fit}}}{\sigma_j}\right )^2\,,
\end{align}
%%%%%%%%%%%%%%%%%%%%%
$\sigma_j$ is the $1\sigma$ error-bar.

% \section{$\chi^2$ mini fitting: Yukawas from Antusch arXiv 2510.01312}
%%%%%%%%%%%%%%%%%%%%%%%%%%%%%%
%
\subsection{Input: ``inclusive'' data  for $\theta_{23}$ and 
$\theta_{13}$ with $\tan \beta=10$ and $M_{\rm SUSY}=10$ TeV}
%
%%%%%%%%%%%%%%%%%%%%%%%%%%

The results of the fit using as input 
the quark mass rations quoted in 
Eqs. (\ref{eq:mumt}), (\ref{eq:mcmt}), (\ref{eq:mdmb}) and 
(\ref{eq:msmb}), as well as 
the ``inclusive'' data on $\theta_{23}$ and 
$\theta_{13}$ for $\tan\beta=10$ and $M_{\rm SUSY}=10$ TeV
are shown in Table \ref{tab:tanb10Msusy10inclusive}.
%%%%%%%%%%%%%%%%%%%%%%%%%%%%%
\begin{table}[h]
	\footnotesize{
		%\label{tab:1}
		\begin{center}
			\renewcommand{\arraystretch}{1.1}
			\begin{tabular}{|c|c|c|c|c|c|c|c|c|} \hline
				\rule[14pt]{0pt}{3pt}  
				& $\frac{m_s}{m_b}\hskip -1 mm\times\hskip -1 mm 10^2$ 
				& $\frac{m_d}{m_b}\hskip -1 mm\times\hskip -1 mm 10^4$& $\frac{m_c}{m_t}\hskip -1 mm\times\hskip -1 mm 10^3$&$\frac{m_u}{m_t}\hskip -1 mm\times\hskip -1 mm 10^6$&
				$\theta_{12}$ &$\theta_{23}$ &$\theta_{13}$& $\delta$
				\\
				\hline
				\rule[14pt]{0pt}{3pt}  
				Fit &$1.81$ & $8.84$
				& $2.86$ & $5.66$&
				$0.225687$ & $0.0393$ & $0.00442$ &
				$65.6^\circ$
				\\ \hline
				\rule[14pt]{0pt}{3pt}
				Exp	 &$1.80$ & $8.99$ 
				& $2.87$& $ 5.67$ &
				$0.226235$ & $0.0393$ & $0.00384$ &$65.7^\circ$\\
				$1\,\sigma$	&$\pm 0.037$ &$\pm 0.277$ & $\pm 0.105$& $\pm 0.229$
				 &$ \pm 0.00085$ &	$ \pm 0.000466$ & $ ^ {+0.0002347}_{-0.0002396}$ &
				$\pm 3.0^\circ$\\ \hline  
				\rule[14pt]{0pt}{3pt}  
				$\Delta\chi^2_i$ &$0.085$ & $0.289$
				& $0.0005$ & $0.0006$&
				$0.415$ & $0.0008$ & $6.202$ &
				$9.7\times 10^{-4}$	
				\\ \hline
			\end{tabular}
		\end{center}
		\caption{Results of the fits of the quark mass ratios, 
			CKM mixing angles and $\delta_{\rm CP}$,
obtained using, in particular, the ``inclusive'' decay data 
on $\theta_{13}$ and $\theta_{23}$. 'Exp' denotes the  values of 
the observables at the GUT scale, including $1\sigma$ error, 
calculated by accounting for the RG effects 
for $\tan\beta = 10$ and $M_{\rm SUSY} = 10$ TeV. 
See text for further details.
		}
		\label{tab:tanb10Msusy10inclusive}
	}
\end{table}
%%%%%%%%%%%%%%%%%%%%%%%%%%%%%%%%%%%%%%
%
\noindent 
With $\Delta\chi^2=6.99$, which is sum of  $\Delta\chi^2_i$
shown in the Table \ref{tab:tanb10Msusy10inclusive}, 
the quality of the fit is sufficiently 
good. 
For the values of the parameters of the model we find:
% \vspace{-0.8cm}
%%%%%%%%%%%%%%%%%%%%%%%%%%
\begin{align}
\frac{\alpha_d}{\gamma_d}=0.144,\qquad  \frac{\beta_d}{\gamma_d}=0.0720,\qquad
\frac{\alpha_u}{\gamma_u}=0.0182,\qquad  \frac{\beta_u}{\gamma_u}=0.00206\,,
\end{align}
%%%%%%%%%%%%%%%%%%%%%%%%%%%%%
%%%%%%%%%%%%%%%%%%%%%%%%
\begin{align}
\tau= i\,2.2775\,,\quad |q^{(4)}|=0.0279456\,,\quad |g_d|=1.305\,,
\quad \phi_{g_d}=305.3^\circ,  \quad  \sum_i \Delta\chi^2_i=6.99\,. 
% \sum \chi^2=6.99\,.
\label{}
\end{align}
%%%%%%%%%%%%%%%%%%%%%%%%%%%%%%%%%%%%%%%
%
We see that the constant ratios in the up-type quark sector exhibit
a larger difference - by a factor of approximately 9,  
than the constant ratios in the down-type quark sector, where they differ by 
a factor of 2. The values of ``hatted'' constants, which are obtained 
directly from the fit, read:
%%%%%%%%%%%%%%%%%%%%%%%%
\begin{align}
\frac{\hat\beta_d}{\hat\alpha_d}=0.110,\qquad  
% \frac{\hat\beta_d}{|g_d|\hat\gamma_d}=0.0515266,\qquad
\frac{\hat\alpha_d}{|g_d|\hat\gamma_d}=0.0515,\qquad
\frac{\hat\alpha_u}{\hat\beta_u}=18.8,\qquad  \frac{\hat\gamma_u}{\hat\beta_u}=2213\,.
\end{align}
%%%%%%%%%%%%%%%%%%%%%%%%%%%%%%%%%%%%%%%
%
%%%%%%%%%%%%%%%%%%%%%%%%%%%%%%
%
\subsection{Input: ``exclusive'' data  for $\theta_{23}$ and 
$\theta_{13}$ with $\tan \beta=10$ and $M_{\rm SUSY}=10$ TeV}
%
%%%%%%%%%%%%%%%%%%%%%%%%%%
%
We get a bad quality of the fit when using as 
input the ``exclusive'' data on $\theta_{23}$ and 
$\theta_{13}$ with $\tan\beta=10$ and $M_{\rm SUSY}=10$ TeV.
The results of the fit are shown in Table \ref{tab:tanb10Msusy10exclusive}.
%%%%%%%%%%%%%%%%%%%%%%%%%%%%%
\begin{table}[h]
\footnotesize{
		%\label{tab:1}
		\begin{center}
			\renewcommand{\arraystretch}{1.1}
			\begin{tabular}{|c|c|c|c|c|c|c|c|c|} \hline
				\rule[14pt]{0pt}{3pt}  
				& $\frac{m_s}{m_b}\hskip -1 mm\times\hskip -1 mm 10^2$ 
				& $\frac{m_d}{m_b}\hskip -1 mm\times\hskip -1 mm 10^4$& $\frac{m_c}{m_t}\hskip -1 mm\times\hskip -1 mm 10^3$&$\frac{m_u}{m_t}\hskip -1 mm\times\hskip -1 mm 10^6$&
				$\theta_{12}$ &$\theta_{23}$ &$\theta_{13}$& $\delta$
				\\
				\hline
				\rule[14pt]{0pt}{3pt}  
				Fit &$1.83$ & $8.62$
				& $2.87$ & $5.68$&
				$0.224755$ & $0.0370$ & $0.00434$ &
				$65.9^\circ$
				\\ \hline
				\rule[14pt]{0pt}{3pt}
				Exp	 &$1.80$ & $8.99$ 
				& $2.87$& $ 5.67$ &
				$0.226235$ & $0.0371$ & $0.00344$ &$65.7^\circ$\\
				$1\,\sigma$	&$\pm 0.037$ &$\pm 0.277$ & $\pm 0.105$& $\pm 0.229$
				&$ \pm 0.00085$ &	$ \pm 0.00056$ & $ \pm 0.000145$ &
				$\pm 3.0^\circ$\\ \hline  
				\rule[14pt]{0pt}{3pt}  
				$\Delta\chi^2_i$ &$0.639$ & $1.802$
				& $0.004$ & $0.001$&
				$3.03$ & $0.015$ & $38.17$ &
				$0.004$
				\\ \hline
			\end{tabular}
		\end{center}
		\caption{The same as in Table \ref{tab:tanb10Msusy10inclusive},
but with the results obtained using the ``exclusive'' decay data on 
$\theta_{13}$ and $\theta_{23}$. See text for further details.
% Results of the fits of the quark mass ratios, 
% CKM mixing angles and $\delta_{\rm CP}$. 'Exp' denotes the  values of 
% the observables at the GUT scale, including $1\sigma$ error.
}
		\label{tab:tanb10Msusy10exclusive}
	}
\end{table}
%%%%%%%%%%%%%%%%%%%%%%%%%%%%%%%%%%%%%%
%
\noindent In this case $ \Delta\chi^2 =43.66$, 
which implies that the model 
is not compatible with 
 these data. The main problem, as follows from the 
results shown in Table \ref{tab:tanb10Msusy10exclusive}, 
is reproducing the ``exclusive'' value of $\theta_{13}$ 
together with the values of 
$\theta_{12}$, $m_d/m_b$ and $m_s/m_b$.
Indeed, as the analytic expression for 
$\theta_{13}$ given in Eq. (\ref{eq:th13}) suggests, one can tune 
the value of the parameter $|q_4|$ to reproduce the experimental value 
of $\theta_{13}$. This would require somewhat smaller value of 
$|q_4|$ than the one found in the fit of the ``inclusive'' data. 
Correspondingly, to get a good quality of the fit of 
$\theta_{12}$, requires, as it follows from Eqs. (\ref{eq:tan2th12d}), 
(\ref{eq:thu12}) and (\ref{eq:th12}), a significantly 
smaller value of the constant ratio $\hat\beta_d/\hat\alpha_d$. 
This results in a bad quality of the fit of the quark 
mass ratio $m_d/m_b$, once the other down-type quark mass ratio 
$m_s/m_b$ is reproduced.

We give for completeness the values of the constants obtained in the fit: 
% \vspace{-0.8cm}
%%%%%%%%%%%%%%%%%%%%%%%%%%
\begin{align}
\frac{\alpha_d}{\gamma_d}=0.153,\qquad  \frac{\beta_d}{\gamma_d}=0.0743,\qquad
\frac{\alpha_u}{\gamma_u}=0.0184,\qquad  \frac{\beta_u}{\gamma_u}=0.00211\,,
\end{align}
%%%%%%%%%%%%%%%%%%%%%%%%%%%%%

%%%%%%%%%%%%%%%%%%%%%%%%
\begin{align}
\tau= i\,2.28182\,,\quad |q^{(4)}|=0.0277567\,,\quad |g_d|=1.360\,,
\quad \phi_{g_d}=305.5^\circ,  
\quad \sum_i   \Delta\chi^2_i=43.66\,.   
% \sum \chi^2=43.67\,.
\label{}
\end{align}
%%%%%%%%%%%%%%%%%%%%%%%%%%%%%%%%%%%%%%%
%
We see that their values do not differ significantly from the values 
found in the ``inclusive'' data fit.
The ratios of  ``hatted'' constants read:
%%%%%%%%%%%%%%%%%%%%%%%%
\begin{align}
\frac{\hat\beta_d}{\hat\alpha_d}=0.107,\qquad  
% \frac{\hat\beta_d}{|g_d|\hat\gamma_d}=0.052736,\qquad
\frac{\hat\alpha_d}{|g_d|\hat\gamma_d}=0.0525\,,\qquad
\frac{\hat\alpha_u}{\hat\beta_u}=18.6,
\qquad  \frac{\hat\gamma_u}{\hat\beta_u}=2163\,.
\end{align}
%%%%%%%%%%%%%%%%%%%%%%%%%%%%%%%%%%%%%%%

%%%%%%%%%%%%%%%%%%%%%%%%%%%%%%
%
\subsection{Input: ``average'' data  for $\theta_{23}$ and 
$\theta_{13}$ with $\tan \beta=10$ and $M_{\rm SUSY}=10$ TeV}
%
%%%%%%%%%%%%%%%%%%%%%%%%%%
%
The results of the fit are presented in Table  \ref{tab:tanb10Msusy10average}.
Although in this case $ \Delta\chi^2 =22.51$ 
and thus is significantly smaller that in the 
``exclusive'' case, it is still quite large showing that 
the model is
%%%%%%%%%%%%%%%%%%%%%%%%%%%%%
\begin{table}[h]
	\footnotesize{
		%\label{tab:1}
		\begin{center}
			\renewcommand{\arraystretch}{1.1}
			\begin{tabular}{|c|c|c|c|c|c|c|c|c|} \hline
				\rule[14pt]{0pt}{3pt}  
				& $\frac{m_s}{m_b}\hskip -1 mm\times\hskip -1 mm 10^2$ 
				& $\frac{m_d}{m_b}\hskip -1 mm\times\hskip -1 mm 10^4$& $\frac{m_c}{m_t}\hskip -1 mm\times\hskip -1 mm 10^3$&$\frac{m_u}{m_t}\hskip -1 mm\times\hskip -1 mm 10^6$&
				$\theta_{12}$ &$\theta_{23}$ &$\theta_{13}$& $\delta$
				\\
				\hline
				\rule[14pt]{0pt}{3pt}  
				Fit &$1.82$ & $8.79$
				& $2.88$ & $5.67$&
				$0.225276$ & $0.0383$ & $0.00439$ &
				$65.8^\circ$
				\\ \hline
				\rule[14pt]{0pt}{3pt}
				Exp	 &$1.80$ & $8.99$ 
				& $2.87$& $ 5.67$ &
				$0.226235$ & $0.0383$ & $0.00355$ &$65.7^\circ$\\
				$1\,\sigma$	&$\pm 0.037$ &$\pm 0.277$ & $\pm 0.105$& $\pm 0.229$
				&$ \pm 0.00085$ &	$ \pm 0.00112$ & $ \pm 0.000186$ &
				$\pm 3.0^\circ$\\ \hline  
				\rule[14pt]{0pt}{3pt}  
				$\Delta\chi^2_i$&$0.294$ & $0.558$
				& $0.009$ & $2.6\times 10^{-4}$&
				$1.27$ & $5.8\times 10^{-6}$ & $20.37$ &
				$0.001$
				\\ \hline
			\end{tabular}
		\end{center}
		\caption{
The same as in Table \ref{tab:tanb10Msusy10inclusive},
but with the results obtained using the ``average'' experimental values of  
$\theta_{13}$ and $\theta_{23}$. See text for further details.
% Results of the fits of the quark mass ratios, 
% CKM mixing angles and $\delta_{\rm CP}$. 'Exp' denotes the  values 
% of the observables at the GUT scale, including $1\sigma$ error.
}
\label{tab:tanb10Msusy10average}
}
\end{table}
%%%%%%%%%%%%%%%%%%%%%%%%%%%%%%%%%%%%%%
% 
\noindent disfavored (if not ruled out) essentially by the 
``average'' data on $\theta_{13}$. The main problem as in the 
``exclusive''  data case is reproducing 
the ``average'' value of $\theta_{13}$ 
together with the values of $\theta_{12}$, $m_d/m_b$ and $m_s/m_b$.
For the values of the parameters of the model we have obtained:

\vspace{-0.3cm}
%%%%%%%%%%%%%%%%%%%%%%%%%%
\begin{align}
\frac{\alpha_d}{\gamma_d}=0.147,\qquad  \frac{\beta_d}{\gamma_d}=0.0733\qquad
\frac{\alpha_u}{\gamma_u}=0.0183,\qquad  \frac{\beta_u}{\gamma_u}=0.00208\,,
\end{align}
%%%%%%%%%%%%%%%%%%%%%%%%%%%%%
%%%%%%%%%%%%%%%%%%%%%%%%
\begin{align}
\tau= i\,2.27919\,,\quad |q^{(4)}|=0.02787\,,\quad |g_d|=1.329\,,\quad \phi_{g_d}=305.3^\circ,  \quad    \sum_i\Delta\chi^2_i = 22.51\,.
\label{}
\end{align}
%%%%%%%%%%%%%%%%%%%%%%%%%%%%%%%%%%%%%%%
%

\vspace{-0.4cm}
We give below also the values of the ``hatted'' constant ratios:  
%%%%%%%%%%%%%%%%%%%%%%%%
\begin{align}
\frac{\hat\beta_d}{\hat\alpha_d}=0.109,\qquad 
% \frac{\hat\beta_d}{|g_d|\hat\gamma_d}=0.0518986,\qquad
\frac{\hat\alpha_d}{|g_d|\hat\gamma_d}=0.0520,\qquad
\frac{\hat\alpha_u}{\hat\beta_u}=18.8,\qquad  \frac{\hat\gamma_u}{\hat\beta_u}=2192\,.
\end{align}
%%%%%%%%%%%%%%%%%%%%%%%%%%%%%%%%%%%%%%%

%%%%%%%%%%%%%%%%%%%%%%%%%%%%%%
%
\subsection{Input: ``average'' data  for $\theta_{23}$ and 
$\theta_{13}$ with $M_{\rm SUSY} = M_{\rm GUT}$ }
%
%%%%%%%%%%%%%%%%%%%%%%%%%%

We get a very good quality of the fit of the 
data in the case of a relatively high SUSY breaking scale.
We present in Table \ref{tab:MsusyEqMgutaverage}
results in the ``extreme'' case of 
$M_{\rm SUSY} = M_{\rm GUT}$, using as input 
the values of the quark mass ratios, of $\theta_{12}$,
the ``average'' experimental values of 
$\theta_{13}$ and $\theta_{23}$
given in Eq. (\ref{eq:atMGUTSM}), and the value of 
$\delta$ quoted in Eq. (\ref{eq:dCPV}) 
($\delta$ practically does not change with the scale). 
As we have already noticed, in this case the 
proper SUSY effects in the RG evolution of 
the observables are negligible and  
the values of the observables evolve with the 
scale as in the Standard Model (SM).  
In contrast to the case of, e.g., $M_{\rm SUSY} = 10$ TeV and $\tan\beta = 10$, 
the RG evolution of the values of $\theta_{13}$ and $\theta_{23}$ 
in the SM is opposite to that for a relatively low value of 
$M_{\rm SUSY}$: $\theta_{13}$ and $\theta_{23}$ increase with the  scale 
instead of decreasing \cite{Antusch:2025fpm}.
As a consequence, describing the values of $\theta_{13}$, 
$\theta_{12}$, $m_d/m_b$ and $m_s/m_b$ at $M_{\rm GUT}$ 
does not represent a problem and we get 
$ \Delta\chi^2 =0.499$. 
%%%%%%%%%%%%%%%%%%%%%%%%%%%%%
\begin{table}[h]
\footnotesize{
		%\label{tab:1}
		\begin{center}
			\renewcommand{\arraystretch}{1.1}
			\begin{tabular}{|c|c|c|c|c|c|c|c|c|} \hline
				\rule[14pt]{0pt}{3pt}  
				& $\frac{m_s}{m_b}\hskip -1 mm\times\hskip -1 mm 10^2$ 
				& $\frac{m_d}{m_b}\hskip -1 mm\times\hskip -1 mm 10^3$& $\frac{m_c}{m_t}\hskip -1 mm\times\hskip -1 mm 10^3$&$\frac{m_u}{m_t}\hskip -1 mm\times\hskip -1 mm 10^6$&
				$\theta_{12}$ &$\theta_{23}$ &$\theta_{13}$& $\delta$
				\\
				\hline
				\rule[14pt]{0pt}{3pt}  
				Fit &$2.14$ & $1.07$
				& $3.25$ & $6.44$&
				$0.225994$ & $0.0465$ & $0.00446$ &
				$65.7^\circ$
				\\ \hline
				\rule[14pt]{0pt}{3pt}
				Exp	 &$2.13$ & $1.07$ 
				& $3.26$& $ 6.44$ &
				$0.226235$ & $0.0465$ & $0.00433$ &$65.
				7^\circ$\\
				$1\,\sigma$	&$\pm 0.054$ &$\pm 0.27$ & $\pm 0.10$& $\pm 0.227$
				&$ \pm 0.00085$ &	$ \pm 0.00136$ & $ \pm 0.00023$ &
				$\pm 3.0^\circ$\\ \hline  
				\rule[14pt]{0pt}{3pt}  
				$\Delta\chi^2_i$&$0.033$ & $0.046$
				& $0.003$ & $1.7\times 10^{-4}$&
				$0.080$ &$2.6\times 10^{-4}$& $0.336$ &
				$2.6\times 10^{-4}$
				\\ \hline
			\end{tabular}
		\end{center}
\caption{Results of the fits of the quark mass ratios, 
CKM mixing angles and $\delta_{\rm CP}$, using as input, in particular, 
the ``average'' experimental values of $\theta_{13}$ and $\theta_{23}$.
'Exp' denotes the  values of the observables at the GUT scale, 
including $1\sigma$ error, calculated by accounting for the RG effects 
in the case of  $M_{\rm SUSY} = M_{\rm GUT}$. See text for further details.
}
\label{tab:MsusyEqMgutaverage}
}
\end{table}
%%%%%%%%%%%%%%%%%%%%%%%%%%%%%%%%%%%%%%

% \vspace{-0.3cm}
For the values of the ratios of the constant parameters we get from the fit:
%%%%%%%%%%%%%%%%%%%%%%%%%%
\begin{align}
\frac{\alpha_d}{\gamma_d}=0.146,\qquad  \frac{\beta_d}{\gamma_d}=0.0747\qquad
\frac{\alpha_u}{\gamma_u}=0.0206,\qquad  \frac{\beta_u}{\gamma_u}=0.00232\,,
\end{align}
%%%%%%%%%%%%%%%%%%%%%%%%%%%%%

\vspace{-0.6cm}
%%%%%%%%%%%%%%%%%%%%%%%%
\begin{align}
\tau= i\,2.2758\,,\quad |q^{(4)}|=0.0280203\,,\quad |g_d|=1.128\,,\quad \phi_{g_d}=303.8^\circ,  \quad  
 \sum_i \Delta\chi^2_i=0.499\,.
\label{}
\end{align}
%%%%%%%%%%%%%%%%%%%%%%%%%%%%%%%%%%%%%%%
%

\vspace{-0.4cm}
For the ratios of the ``hatted'' constants we have obtained:
%%%%%%%%%%%%%%%%%%%%%%%%
\begin{align}
\frac{\hat\beta_d}{\hat\alpha_d}=0.112,\qquad  
% \frac{\hat\beta_d}{|g_d|\hat\gamma_d}=0.06078,\qquad
\frac{\hat\alpha_d}{|g_d|\hat\gamma_d}=0.0607,\qquad
\frac{\hat\alpha_u}{\hat\beta_u}=18.9,\qquad  \frac{\hat\gamma_u}{\hat\beta_u}=1962\,.
\end{align}
%%%%%%%%%%%%%%%%%%%%%%%%%%%%%%%%%%%%%%%
%

 Given the fact that the ``inclusive'' decay data best fit values of 
$\theta_{13}$ and $\theta_{23}$ are larger and have larger 
uncertainties than the corresponding ``average'' values, 
the model in the case of $M_{\rm SUSY} \sim M_{\rm GUT}$ 
provides a good quality fit of the 
quark observables also if one uses as input the 
 ``inclusive'' decay data on $\theta_{13}$ and $\theta_{23}$.

\vspace{0.6cm}
Finally, we comment on the effect of  non-vanishing  real $\tau$,
$\re\tau=[-\frac12,+\frac12]$, 
 on our results.
When taken into account, 
the numerical result on the mixing angles, CP phase and quark 
mass ratios change by less than $0.5\%$ in the considered four 
cases of input data.

%%%%%%%%%%%%%%%%%%%%%%%%%%%%%%
%
\section{Summary and Conclusions}
\label{sec:summary}
%
%%%%%%%%%%%%%%%%%%%%%%%%%%

In the present article we have addressed the problem present in
``bottom-up'' modular invariant quark flavour models of 
reconciling the non-fine-tuned generation of the  quark mass hierarchies 
with a correct description of quark mixing and CP-violation,  
when the only source of CP and flavour symmetries breaking  
is the VEV of a single modulus $\tau$, $\tau_\text{vev}$.
We have constructed a models with 
$S^\prime_4$ modular (flavour) symmetry with the requirements 
of being phenomenologically viable and having a minimal 
number of parameters.
The quark doublet superfields $Q$ are assumed in the model 
to furnish the triplet representation $\mathbf{3}$
of the finite modular group $S^\prime_4$,  
% and have a weight $k_Q$, 
while the RH up-type quark  superfields $u^c$, $c^c$ and $t^c$ 
and  RH down-type quark  
superfields $d^c$, $s^c$ and $b^c$ 
furnish the singlet representations 
$\mathbf{1}$, $\mathbf{\hat{1}}$, $\mathbf{\hat{1}}$
and $\mathbf{1}$, $\mathbf{1}$, $\mathbf{\hat{1}}$, respectively. 
% 
% and have weights $k_{u^c}$, $k_{c^c}$,$k_{t^c}$ and 
% $k_{d^c}$, $k_{s^c}$,$k_{b^c}$.
% We assume further that $k_Q + k_{u^c} = 4$, 
% $k_Q + k_{c^c} = 3$, $k_Q + k_{t^c} = 5$,
% $k_Q + k_{d^c} = 6$, 
% $k_Q + k_{s^c} = 4$ and $k_Q + k_{b^c} = 7$.
Assigning specific modular weights to 
$Q$, $u^c$, $c^c$ and $t^c$,  $d^c$, $s^c$ and $b^c$,
leads, as a consequence of the  modular invariance, 
to up-type quark mass matrix ($M_u$)
involving three modular forms of weights 4, 3 and 5,   
which transform as  $\mathbf{3}$, $\mathbf{\hat{3}'}$
and $\mathbf{\hat{3}'}$ representations of $S^\prime_4$,
and three constant parameters,
and to down-type quark mass matrix ($M_d$)
formed by two triplet ($\mathbf{3}$)
modular forms of weights 
6 and 4, and two modular forms of weight 7 
furnishing the same $\mathbf{\hat{3}'}$
representation of  $S^\prime_4$, 
and, correspondingly, including four 
constant parameters. 
The three constant parameters in $M_u$
% the up-type quark mass matrix,  
and three of the four constant parameters 
% in the down-type quark mass matrix, 
in $M_d$ can be 
taken to be real without loss of generality.
The fourth parameter 
% in the   down-type quark mass matrix,
in $M_d$, however, will, in general, be complex and 
CP-violating if the gCP-symmetry is not imposed.

 In the approach  we followed in the present study 
the quark mass hierarchies are considered in the ``vicinity'' of the 
fixed point $\tau_\text{T} = i\infty$ of the 
(full homogeneous) modular group, preserving the 
$\mathbb{Z}^T_4$ symmetry, $T$ being one of the generators of the 
finite modular group  $S^\prime_4$ 
(Sections \ref{sec:Modular} and \ref{sec:theory}). 
The deviation of the VEV of the modulus 
$\tau_\text{vev}$ from $\tau_\text{T}$
serves as a small parameter in our analysis. 
In the case we consider this deviation is described 
by $|q_4| = \exp(-\,2\pi\,\im [\tau_\text{vev}]/4) \ll 1$.
Owing to the behavior of the modular forms 
present in the quark mass matrices 
in the vicinity of $\tau_\text{T}$,  
the elements of the mass matrices 
are expressed as powers 
of the small parameter $|q_4|$, the exponents of which 
can take values $l=0,1,2,3$ and  
are uniquely determined by the transformation properties 
of the respective quark field components 
under the action of the residual symmetry group $\mathbb{Z}^T_4$.
Thus, hierarchical quark mass matrices are generated leading to a 
non-fine-tuned  generation of the quark mass hierarchies. 

When implemented in ``bottom-up'' minimal modular quark flavour models, 
characterised by CP-conserving constant parameters and 
only one modulus $\tau$ whose VEV is the only source of 
CP-symmetry breaking, the described approach employed by us  
faces the problem of reconciling the relatively small values of the 
non-diagonal elements of the quark mass matrices, needed to generate 
the quark mass hierarchies, with reproducing the relatively large 
value of the CPV phase in the CKM mixing matrix. 
To avoid this problem  
in the quark flavour model constructed by us,  
the CP-symmetry is broken explicitly by the complex constant 
parameter present naturally, i.e.,  without additional assumptions,
in the down-type quark mass matrix $M_d$ of the model, as 
the gCP symmetry is not imposed. 
The $\re [\tau_\text{vev}]$ can be non-zero, implying spontaneous 
violation of the gCP symmetry, but its effects are so strongly suppressed 
that it plays no role not only in the generation of the observed 
CP-violation in the quark sector, 
but also in the description of the quark  mixing. 
Thus effectively, the ten quark observables are described 
by nine real parameters in the model.

As a consequence of the relatively fastly converging 
$q_4$-expansions of the modular forms present in the quark mass 
matrices, we were able to solve the model completely  analytically, 
i.e., to derived  sufficiently precise analytic expressions for all 
ten observables  in terms of the eight real constant parameters of the model 
and $\im [\tau_\text{vev}]$ (or $|q_4|$). 
These expressions turned out to be surprisingly simple. 
We further performed statistical analyses of the model assuming 
i) relatively low SUSY breaking scale $M_{\rm SUSY} = 10$ TeV, 
$\tan\beta = 10$, and using, in particular, 
the ``inclusive'' and ``exclusive'' decay data on,  
and the ``average'' experimental values of,  
the $|V_{ub}|$ and $|V_{cb}|$ elements of the CKM matrix,
as well as ii) the ``average'' experimental values of 
the $|V_{ub}|$ and $|V_{cb}|$ in the case of very high $M_{\rm SUSY}$. 
The results of these analyses showed  
that the quark flavour model with $S^\prime_4$ 
modular symmetry constructed by us,  
is phenomenologically viable and 
consistent, in particular, with the ``inclusive'' decay data 
on the $|V_{ub}|$ and $|V_{cb}|$ elements of the CKM matrix and,
in the case of a very high scale of supersymmetry breaking, 
with the current ``average'' experimental values 
of $|V_{ub}|$ and $|V_{cb}|$.

Our study is part of the ongoing quest of finding 
"The Standard Model of Flavour (SMF)".   
The results we have obtained, 
together with the results of other authors 
indicate that the modular invariance 
is a promising approach to this fundamental problem
and should be further explored.

\newpage
\small
%============
\paragraph{Acknowledgements:}
%============
M.T is grateful to Zhi-zhong Xing and Shun Zhou for useful discussions.
S.T.P. would like to thank S. Antusch and M. Spinrath for discussions.
The work of S.T.P. was supported in part by 
the European Union's Horizon Europe research and innovation programme 
under the Marie Sk\l{}odowska-Curie Staff Exchange grant 
agreement No.~101086085-ASYMMETRY,  
by the Italian INFN program on Theoretical Astroparticle Physics and by 
the World Premier International Research Center Initiative 
(WPI Initiative, MEXT), Japan.

\end{document}